\documentclass[12pt]{article}

\pdfoutput=1

\usepackage{amsmath,amssymb,bbm}
\usepackage{epsfig}
\usepackage{slashed}
\usepackage{color}

\def\Dslash{\hspace{3pt}\raisebox{1pt}{$\slash$} \hspace{-9pt} D}

\title{
\vspace{-2cm}
\begin{flushright}
\small{CERN-PH-TH/2012-287}
\end{flushright}
\vspace{3cm}
\bf \huge
On the Tuning and the Mass \\of the Composite Higgs
\vspace{.2cm}}
\date{}
\author{
{\large Giuliano Panico$^{a,b}$\footnote{panico@phys.ethz.ch}, Michele Redi$^{c,a}$\footnote{michele.redi@cern.ch}, Andrea Tesi$^{d}$\footnote{andrea.tesi@sns.it}, Andrea Wulzer$^{e}$\footnote{a.wulzer@gmail.com}}\\
[10mm]
\normalsize\itshape $^a$ CERN, Theory Division, CH-1211, Geneva 23, Switzerland\\
\normalsize\itshape $^b$ Institute for Theoretical Physics, ETH, CH-8093, Zurich, Switzerland\\
\normalsize\itshape $^c$ INFN, Sezione di Firenze, Via G. Sansone, 1; I-50019 Sesto Fiorentino, Italy\\
\normalsize\itshape $^d$ Scuola Normale Superiore, Piazza dei Cavalieri, 7; I-56126 Pisa, Italy\\
\normalsize\itshape $^e$ Dipartimento di Fisica e Astronomia and INFN, Sezione di Padova,\\
\normalsize\itshape via Marzolo 8, I-35131 Padova, Italy
}

\begin{document}
\maketitle
\begin{abstract}
\medskip
\noindent
We analyze quantitatively the tuning of composite Higgs models with partial compositeness and its interplay with the predicted Higgs mass. In this respect we identify three classes of models, characterized by different quantum numbers of the fermionic colored resonances associated with the top quark, the so-called top partners. The main result of this classification is that in all models with moderate tuning a light Higgs, of 125 GeV mass, requires the presence of light top partners, around 1 TeV. The minimal tuning is comparable to the one of the most attractive supersymmetric models in particular the ones realizing Natural SUSY. This gives further support to an extensive program of top partners searches at the LHC that can already probe the natural region of composite Higgs models.
\end{abstract}

\newpage
\section{Introduction}

Despite its many successes, the Standard Model (SM) might not be the complete
description of the Electro-Weak Symmetry Breaking (EWSB). The concept of
``naturalness'', incarnated by the famous Hierarchy Problem, puts strong doubts
on the possibility that an elementary weakly-coupled Higgs doublet is entirely
responsible for EWSB with no new degrees of freedom and interactions appearing
below a very high energy scale such as the Planck mass $M_P$. The problem is that
if the SM emerges from a fundamental theory with typical scale $M_P$ a realistic Higgs
mass term, of the order of $100$~GeV, can only be obtained at a price of an ``unnatural''
cancellation taking place in the Higgs potential.

From the above discussion it is clear why the concept of naturalness is so important in
any extension of the SM that aims to solve the Hierarchy Problem. Any such model is
characterized by a
new physics scale $\Lambda$ which is typically constrained by observations to lie in the
TeV or multi-TeV region, well above the EWSB scale. A certain amount of tuning, $\Delta$,
has to be performed on the parameters of the model in order to achieve this separation
of scales, but, if naturalness is really relevant for Nature, the tuning must be reasonably small.

The recent observation of a Higgs-like particle around $125$~GeV \cite{:2012gk,:2012gu} adds
a new relevant piece of information to this discussion. Accommodating the observed
Higgs mass into a model might be problematic and require additional tuning. For
instance, in the MSSM the Higgs would naturally be lighter, raising it to the observed
value requires increasing the stop mass worsening the level of tuning. In the case of
composite Higgs models (CHM), which we consider in the present paper, the situation is
basically reversed. As we will see, models with moderate tuning typically predict
a too heavy Higgs. A realistic Higgs mass can be incorporated either at the price
of fine-tuning, similarly to the MSSM, or by lowering the mass of the fermionic resonances
associated to the top quark, the so-called ``top-partners''.

The aim of this paper is to study quantitatively the fine-tuning issue in the context of composite-Higgs models with partial fermion compositeness~\cite{pc} and to analyze the interplay among the
tuning and the Higgs mass.  In the scenario we consider the Higgs is a pseudo
Nambu-Goldstone
boson (pNGB) associated with a spontaneous global symmetry breaking, most simply
\mbox{SO$(5)$ $\to$ SO$(4)$}~\cite{gk,Agashe:2004rs,Contino:2006qr} (see also~\cite{DP}).
The pNGB Higgs is characterized by the scale of symmetry breaking $f$ that controls its interactions, in particular the potential.
Various experimental results require that the electroweak-scale $v\simeq\langle h\rangle < f$.
This can be achieved through cancellations in the potential with a precision that scales as $\Delta=f^2/v^2$.
A tuning of order 10\% is often believed sufficient to comply with the experimental constraints.

However it has been found that the tuning is typically larger in concrete constructions.
This has been verified explicitly in 5d holographic models~\cite{Panico:2006em,Panico:2008bx,Csaki:2008zd}
and explained parametrically in Ref.~\cite{cthdm,Matsedonskyi:2012ym}.
The point is that in specific models the Higgs potential might assume a
non-generic form that renders the cancellation more difficult to realize.
Roughly speaking, it can happen that the mass term is enhanced with respect to the quartic Higgs coupling
so that obtaining a small VEV requires more tuning.
This crucially depends on the structure of the potential that in turn is controlled by
the quantum numbers of the composite fermions.

In the present paper we explore different choices of the fermion representations and classify them
in terms of the structure of the Higgs potential they induce. We find three categories. The first
one is characterized by an enhanced tuning (or ``double tuning'' as we dub it), due to the mechanism described above.
Interestingly enough the most studied models, namely the MCHM$_4$~\cite{Agashe:2004rs}, MCHM$_5$ and MCHM$_{10}$~\cite{Contino:2006qr}, all belong to this category.
The models in the second category are less tuned, to obtain a given scale separation $v/f$
the tuning is $\Delta \sim f^2/v^2$, that is it follows the naive estimate.
We denote this as ``minimal'' tuning because we believe it is impossible to reduce it further
with a better construction. In the third category there are models with minimal tuning
in which the $t_R$ quark is a completely composite state.

The expected size of the Higgs mass $m_h$ is rather different in the three categories
described above. However we find that the only way to obtain a light Higgs with moderate
tuning in any of the three cases is to assume a low enough scale $\Lambda$ for the top partners.
This result is not surprising as the role of the resonances is to cut off the quadratically divergent contribution
to $m_h$ from the top quark loop in the SM. Following Ref.~\cite{Barbieri:2000gf} we have
\begin{equation}
\delta m_h^2=\frac{3}{\sqrt{2}\pi^2}G_Fm_t^2\Lambda^2\;\;
\Rightarrow\;\;\Delta\geq\frac{\delta m_h^2}{m_h^2}=
\left(\frac{\Lambda}{400\,{\textrm{GeV}}}\right)^2\left(\frac{125\,{\textrm{GeV}}}{m_h}\right)^2\,.
\label{barbieri}
\end{equation}
In agreement with the above equation we find that a moderate tuning, $\Delta\simeq10$,
requires new fermionic resonances around $1$~TeV. This bound should be compared
with the one on spin-one ``$\rho$'' resonances. For example from S-parameter estimates
one finds for the electro-weak resonances that $m_\rho\gtrsim2.5$~TeV.
Therefore our result requires a certain separation among the fermionic and bosonic resonance scales.

Obviously the one of eq.~(\ref{barbieri}) is only a lower bound on the tuning.
Low-energy arguments cannot exclude that additional contributions to $\delta m_h^2$ could
be present in the complete theory, worsening the cancellation regardless of the presence of light
fermions. This situation occurs in the popular MCHM$_{4,5,10}$. In that case the
role of the light top partners is to reduce the Higgs quartic coupling,
allowing for a light physical Higgs.
The light resonances do not saturate the
quadratic divergence of $\delta m_h^2$ and the tuning remains large. If we accept a large
tuning there is no reason, a priori, why the fermionic resonances should be light, the correlation
among light Higgs and light resonances could be a peculiarity of the models with doubly tuned
potential. We construct an explicit model, with totally composite $t_R$, where
all the resonances can be heavy and the Higgs mass is reduced by the tuning.

The paper is organized as follows. In section 2 we introduce the framework and discuss
the structure of the Higgs potential. We employ an effective field theory
methodology
mainly based on Ref.~\cite{SILH}. However we introduce a slight conceptual
deformation of the original approach: in order to account for a possible separation among
fermionic and vectorial resonances we assume that two separate mass scales, $m_\psi$
and $m_\rho$, are present in the strong sector. Correspondingly we have two couplings,
$g_\psi=m_\psi/f$ and $g_\rho=m_\rho/f$, contrary to Ref.~\cite{SILH} where a single
coupling $g_\rho$ is present. In addition we also consider the limit of weak strong-sector
coupling $g_\psi\sim g_{SM}\sim1$. In section~3 we introduce our models and estimate
parametrically the amount of fine-tuning and the expected size of the Higgs mass in each
case. In section~4 we go beyond the estimates and analyze concrete calculable
implementations of the models based on the 4d frameworks developed
in Refs.~\cite{Panico:2011pw} and \cite{DeCurtis:2011yx}.
We are then able to compute explicitly the Higgs mass and the tuning, quantifying
the deviations from the parametric estimates. In section~5 we summarize our results
in view of the observed Higgs mass $m_h=125$~GeV, by discussing the model-building
options that can lead to a realistic value. Finally we present our conclusions.

\section{General Structure}

We start by giving a lightning review of the basic ingredients of modern constructions of composite Higgs models with partial compositeness
and introducing our notation. The Higgs is a Goldstone boson arising from the spontaneous symmetry breaking of $G\to H$.
We focus on  the minimal coset $\textrm{SO}(5)/\textrm{SO}(4)$ in this paper but the analysis can be readily generalized. The SM gauge bosons are introduced
as elementary fields, external to the strong sector, and gauge the SM subgroup
of $\textrm{SO}(5)$.~\footnote{Notice that in order to accommodate the correct fermion
hypercharges, an extra $U(1)_X$ global symmetry is needed. The presence of this extra symmetry
does not modify the general discussion of sections~2 and 3, so we will neglect it
for simplicity. We will however include the complete symmetry structure in the explicit
models studied in sections~4 and 5. See Refs.~\cite{Panico:2011pw,DeCurtis:2011yx}
for further details.}
As such, they are coupled linearly to the corresponding
currents and the elementary-composite gauge interactions take the form
\begin{equation}
{\mathcal{L}}_{gauge}=g\,W_\mu J^\mu\,.
\label{gaugem}
\end{equation}
The situation is assumed to be analogous for the SM fermions. They are introduced as elementary
fields and coupled linearly to strong-sector fermionic operators with
equal quantum numbers under the SM
\begin{equation}
{\mathcal{L}}_{fermion}=y_L^f f_L {\mathcal{O}}_L+y_R^f f_R {\mathcal{O}}_R\,.
\label{ferm}
\end{equation}
In the IR, where the strong sector is assumed to confine, the interactions above will give rise to mixings of the elementary degrees of freedom
($W$ and $f_{L,R}$) and the strong sector's resonances associated to the operators $J$ and ${\mathcal{O}}$.
This mechanism realizes the paradigm of partial compositeness, according to which the SM particles ({\it{i.e.}} the mass eigenstates) are a mixture of elementary and composite state.  The analogous phenomenon in QCD is the well known photon-$\rho$ mixing.

The fermion interactions of eq.~(\ref{ferm}) introduce an extra model-building ambiguity besides the choice of the coset $G/H$. One must specify the representations of $G$ in which the fermionic operators ${\mathcal{O}}_{L,R}$ transform, popular possibilities are ${\mathbf{r}}_L={\mathbf{r}}_R={\mathbf{4}},\,{\mathbf{5}}$ or ${\mathbf{10}}$ and correspond respectively to the holographic MCHM$_4$, MCHM$_5$ and MCHM$_{10}$ 5d models. The choice of the representations has a strong impact on the structure of the Higgs potential, controlling for example the Higgs mass.
As it turns out, the models considered in the holographic MCHM all fall in the same universality class for what the Higgs potential is concerned, we thus find particularly important to study alternatives.

Another interesting possibility is that the right-handed $t_R$ quark emerges directly from the
strong sector as a composite chiral state. In this case there is no elementary $t_R$ field and
no $y_R^t$ mixing is present in  eq.~(\ref{ferm}).

\subsection{Split Strongly Interacting Light Higgs}

In order to discuss the implications of the above scenario we need a parametrization of the dynamics of the strong sector.
One can build explicit models, as we will do in the following section, but also rely on model-independent estimates based on generic assumptions on the strong sector along the lines of Ref.~\cite{SILH}. At the simplest possible level the strong sector can be characterized by one scale of confinement, $m_\rho$, corresponding to the lightest vector resonances, and one coupling $g_\rho$, possibly related to the number of colors in a QCD-like theory.
The decay constant of the pNGB Higgs satisfies,
\begin{equation}
m_\rho = g_\rho f\,,
\end{equation}
and the effective action is determined, in absence of unnatural cancellations, by simple power counting rules. The vector resonances contribute at tree-level to the $S$ parameter of Electro-Weak Precision Tests (EWPT)
and for this reason their mass is constrained to the multi-TeV range, $m_\rho\gtrsim3$~TeV being a
relatively safe choice.
Therefore the coupling $g_\rho$ is preferentially large because this allows to decouple the vectors without raising $f$, which would require more fine-tuning. Indeed
we can hope to build a reasonably natural theory, as we will discuss below, only for $f\lesssim1$~TeV which implies $g_\rho\gtrsim3$.

We find it necessary to extend the framework of Ref.~\cite{SILH} by introducing a different scale for the fermionic resonances, or at least for the ones associated to the top quark, {\it{i.e.}} the
top partners. The typical mass of the top partners is denoted as $m_\psi$, the associated
coupling $g_\psi$ is defined by
\begin{equation}
m_\psi = g_\psi f\,.
\end{equation}
We will see that taking $m_\psi< m_\rho$ is practically mandatory to obtain a light Higgs with a mild tuning of the parameters. Importantly,
in the 5d holographic models the fermion masses and couplings are tied to the ones of the vectors because they both originate geometrically from the size of the extra dimension. As such, $g_\psi<g_\rho$ is difficult to realize in 5d constructions. A simple way to construct explicit models implementing this scenario is to employ the more general 4d constructions of Refs.~\cite{Panico:2011pw,DeCurtis:2011yx}.

Within the hypothesis of partial compositeness  the couplings $g$, $y_L$ and $y_R$ are responsible for the generation of all the interactions among the elementary states and the composite Higgs. In particular the SM Yukawas, at leading order in $y_{L,R}$, take the form
\begin{equation}
y_{SM}\simeq \frac{y_Ly_R}{g_\psi}\simeq \epsilon_L\cdot g_\psi \cdot   \epsilon_R\,,
\label{topyukawa}
\end{equation}
where we have introduced the mixings $\epsilon_{L,R}=y_{L,R}/g_\psi$ of the left and right chiralities of SM quarks.
In general the quantities above are matrices in flavor space but in this paper only the third generation
will be relevant. There are few caveats with the above formula that should be kept in mind. First of all it is valid only in an expansion
in the mixings, $\epsilon_{L,R}<1$. Therefore it will be in practice insufficient for the top quark when we will consider the case $g_\psi\sim y_{L,R}$
favored by a light Higgs. However, even if it can be violated at ${\mathcal{O}}(1)$, the formula will still provide a valid parametric estimate. Second,
the formula is parametrically violated if some of the top partners, with specific quantum
numbers, are accidentally lighter than the others \cite{Matsedonskyi:2012ym}.
We will give below, in eq.~(\ref{ytlp}), the correct formula for this case.
Notice that it is essential to include properly the effect of the light top partners in order to understand how a light Higgs can be obtained (at the price of tuning, however) in the 5d holographic models where $g_\psi\simeq g_\rho$. Finally, notice that eq.~(\ref{topyukawa}) also holds in the case of total $t_R$ compositeness if setting $\epsilon_R=1$. In this case one simply finds $y_t\simeq y_L$.

\subsection{Higgs potential}

Loops of elementary fields generate a potential for the Higgs because the elementary-com\-po\-si\-te interactions of eqs.~(\ref{gaugem}) and (\ref{ferm})
break explicitly the $\textrm{SO}(5)$ global symmetry. While this radiative contribution to the potential is unavoidable, other contributions may also exist, for example explicit symmetry breaking effects in the strong sector analogous to quark masses in QCD.
These would not change the analysis substantially and we will neglect them here.
The largest contributions to the potential are typically associated to the largest couplings in the SM,
the top Yukawa and the gauge couplings
\begin{equation}
V(h)= V(h)_{\rm top}+V(h)_{\rm gauge} .
\end{equation}
The gauge contribution would be sub-leading for $g_\rho \sim g_\psi$, and for this reason it is often ignored.
Nevertheless we will include it in what follows because it can be of similar size as the fermion contribution
or even dominant in the  region $g_\psi < g_\rho$ preferred by a light Higgs. Similarly the bottom right quark contribution
could  also be relevant if the fermionic coupling in the bottom sector is large, $g_\psi^{\textrm{bottom}}\sim g_\rho>g_\psi$.
A large coupling in the bottom sector is suggested by EWPT, and in particular by the need of keeping under control the
tree-level corrections to the $Zb_L{\overline{b}}_L$ vertex.

In an expansion in the elementary-composite interactions the Higgs potential is strongly constrained by the $\textrm{SO}(5)$ symmetry.
This is best understood by promoting $g,\,y_L$ and $y_R$ to spurions and noticing that the potential must formally respect the
$\textrm{SO}(5)$ symmetry under which both the Higgs and the spurions
transform~\cite{cthdm,Panico:2011pw}. With this technique it is possible to establish, order by
order in the number of spurion insertions, the functional form of the Higgs potential. Making also use of naive power counting to estimate
the overall size one finds, for the gauge contribution
\begin{equation}
\label{explicit-pot-gauge}
V(h)_{\rm gauge}\sim \frac{9\,g^2}{64\pi^2} \frac {m_{\rho}^4}{g_\rho^2} s_h^2\,,
\end{equation}
where $s_h=\sin h/f$. Higher order terms in the spurion expansion are small, being suppressed by $(g/g_\rho)^2$. Notice that
$V_{\rm gauge}$ is rather model independent
because the quantum numbers under $\textrm{SO}(5)$ of the $g$ spurion in eq.~(\ref{gaugem}) are fixed.

The fermionic contribution, on the contrary, is not universal because the quantum numbers of $y_{L,R}$ depend
on the representation of the fermionic operators ${\mathcal{O}}_{L,R}$.
Once the choice of representations is made, the classification of the invariants can be
carried out in the same way as for the gauge fields. We can obtain the same result in a somewhat more explicit way by first writing down the
effective action for the elementary quarks obtained by integrating out the strong sector, and afterwards computing the Coleman-Weinberg one-loop Higgs potential.
Neglecting higher derivative terms, the structure of the effective Lagrangian obtained integrating out the heavy fermions is schematically,
\begin{eqnarray}
{\cal L}&=&\left(1+ \epsilon_L^2 \sum_i a_i\, f_i(h/f)\right) \, \bar {q}_L  \slashed{D} q_L+ \left(1+ \epsilon_R^2 \sum_i b_i\, g_i(h/f)\right)\, \bar {q}_R  \slashed{D} q_R\nonumber \\
&&+\ y_t f\, \left(\sum_i c_i\, m_i(h/f)\right)  \bar q_L  q_R+ h.c.\,,
\label{bah}
\end{eqnarray}
where the functions $f_i,g_i$ and $m_i$ are trigonometric polynomials  (respectively even, $f_i$ and $g_i$, and odd, $m_i$) determined  by the spurionic analysis for each given choice of the fermion representation. In practice, the number of allowed polynomials is extremely limited in concrete models. As explained in Ref.~\cite{cthdm} the number of say LL invariants corresponds to the number of singlets under $\textrm{SO}(4)$ contained in the product of ${\mathbf{r}}_L \times {\mathbf{r}}_L$ minus one,
where $r_L$ is the $\textrm{SO}(5)$ representation. For instance in the case of the MCHM$_5$ there is only one $f_1=s_h^2$, one $g_1=s_h^2$ and $m_1=\sin(2h/f)$.
The coefficients $a_i$, $b_i$ and $c_i$ are a priori of order one but their values can be reduced either by  tuning or if the fermionic
sector respects some (approximate) global symmetry. We will give an example of this below.

The loops of SM fermions are UV divergent within the low energy theory described by eq.~(\ref{bah}),
but they are cut-off by the non-local form factors which account for the presence of the fermionic resonances of the full theory. The cutoff scale is provided by the scale $m_\psi$ of the fermionic resonances and therefore the Higgs potential takes the form
\begin{eqnarray}
V_{\rm leading} &\sim &\frac {N_c}{16\pi^2}m_\psi^4\sum_i \left[\epsilon_L^2 \,I_L^{(i)}(h/f) + \epsilon_R^2 \,I_R^{(i)}(h/f)\right]\,,\,\nonumber \\
V_{\rm sub-leading} &\sim& \frac {N_c}{16\pi^2} m_\psi^4\, \sum_i \left[
\frac {y_t^2}{g_{\psi}^2}\,  I_{LR}^{(i)}(h/f) +\epsilon_L^4 I_{LL}^{(i)}(h/f)+\epsilon_R^4 I_{RR}^{(i)}(h/f)
\right]
\,.
\label{potential}
\end{eqnarray}
Notice that the term proportional to $y_t^2$ in the above equation is of order $\epsilon_L^2\epsilon_R^2$ (see eq.~(\ref{topyukawa})),
{\it{i.e.}} of the same order as the $\epsilon_{L,R}^4$ ones. The origin of the invariant trigonometric polynomials $I^{(i)}$ can be tracked back to the $f_i,g_i$ and $m_i$
of eq.~(\ref{bah}), and again their number is quite limited in explicit models. In the case of the MCHM$_5$ there is only one quadratic invariant, $I_L=I_R=\sin^2(h/f)$, and
a second one only emerges at the quartic order, $I_{LL} = \sin^2(2h/f)$. The invariants are listed in table~\ref{invariants} for the various cases considered in the present
paper.

\begin{table}[t]
\begin{center}
\begin{tabular}{|c|c|c|}
\hline
\raisebox{-0.5em}{\rule{0pt}{1.5em}}\  & $I_L,\,I_R$ & $I_{LL},\,I_{RR},\,I_{LR}$\\
\hline\hline
\raisebox{-0.5em}{\rule{0pt}{1.5em}}
${\mathbf{r_L}}={\mathbf{r_R}}={\mathbf{5}}$ & $\sin^2({h/f})$ & $\sin^{2n}({h/f})$ \; {\textrm{with}}\; $n=1,2$
\\
\hline
\raisebox{-0.5em}{\rule{0pt}{1.5em}}
${\mathbf{r_L}}={\mathbf{r_R}}={\mathbf{10}}$ & $\sin^2({h/f})$ & $\sin^{2n}({h/f})$ \; {\textrm{with}}\; $n=1,2$
\\
\hline
\raisebox{-0.5em}{\rule{0pt}{1.5em}}
${\mathbf{r_L}}={\mathbf{r_R}}={\mathbf{14}}$ & $\sin^2({h/f})$,\;$\sin^4({h/f})$ & $\sin^{2n}({h/f})$ \; {\textrm{with}}\; $n=1,2,3,4$\\
\hline
\raisebox{-0.5em}{\rule{0pt}{1.5em}}
${\mathbf{r_L}}={\mathbf{r_R}}={\mathbf{4}}$ & $\sin^2({h/2f})$ & $\sin^{2n}(h/2f)$ \; {\textrm{with}}\; $n=1,2$\\
\hline
\end{tabular}
\end{center}
\caption{\label{invariants}\small Table with all possible invariants appearing in the Higgs potential. For the case with totally composite $t_R$ only the $I_L$ and $I_{LL}$ invariants
are relevant.}
\end{table}

One caveat to eq.~(\ref{potential}) is that in the limit of full compositeness, $\epsilon_R\sim 1$ for the top right, there are no contributions in
$\epsilon_R^2$ or $\epsilon_R^4$ because the state is part of the strong sector respecting the global symmetries. In this case the $y_t^2$ term
in the second line of eq.~(\ref{topyukawa}) becomes of the same order of the formally leading $\epsilon_L^2$ because, as mentioned above, $y_L$
becomes of order $y_t$. Indeed in the case of total $t_R$ compositeness there is a single source of breaking of global symmetries, the mixing of the left doublet.
Therefore the expansion is truly in $\epsilon_L^2$. Another important remark is that the very notion of leading and subleading terms becomes useless in the limit
of very small fermionic coupling, $g_\psi\sim y_{L,R}$ because the expansion in $\epsilon_{L,R}$ looses its validity.
In this case, similarly to what we mentioned below eq.~(\ref{topyukawa}) concerning the estimate of the Yukawa couplings, eq.~(\ref{potential}) can be violated at ${\mathcal{O}}(1)$ but still it provides a valid estimate of the size of the Higgs potential.

\section{Tuning and Mass of the Composite Higgs}
\label{sec:2a}

The Higgs potential in eq.~(\ref{potential}) generically has its minimum for $\langle h\rangle\sim f$. The phenomenological
success of the model requires instead $\langle h\rangle<f$, {\it{i.e.}} that the parameter
\begin{equation}
\xi\,=\,\left(\frac{v}{f}\right)^2\,=\,\sin^2{\frac{\langle h\rangle}{f}}\,,
\end{equation}
is smaller than one. As a benchmark in this paper we will mainly focus on the relatively conservative choice $\xi=0.1$, which corresponds to $f\simeq800$~GeV.
Achieving this requires unavoidably
some cancellation. However the actual level of fine-tuning $\Delta$ which has to be enforced crucially depends on the structure of the Higgs potential,
which in turn is determined by the choice of the fermionic representation and also by the size of the fermionic coupling $g_\psi$. For what concerns the
fine-tuning issue the composite Higgs models are conveniently classified into three categories, which we will describe below. The popular MCHM$_4$,
MCHM$_5$ and  MCHM$_{10}$ all belong to the first class and they suffer of an enhanced (or ``double'') amount of tuning. The tuning will be smaller
in the other two categories, it will be of order
\begin{equation}
\Delta=\Delta_{\rm min}=\frac{1}{\xi}\,.
\end{equation}
We refer to $\Delta_{\rm min}=1/\xi$ as the ``minimal tuning'' because we expect that it provides the absolute lower bound for the tuning required by any model
of composite Higgs, for sure this is the case for all the models of the present paper.

\subsection{Double Tuning}

As exhaustively discussed in Ref.~\cite{Matsedonskyi:2012ym}, a parametrically enhanced fine-tuning is needed in all the models where a single invariant is present in
the potential at the leading order in $\epsilon_{L,R}$. In this case the subleading terms must be taken into account in order to achieve a realistic EWSB.
For instance for ${\mathbf{r_L}}={\mathbf{r_R}}={\mathbf{5}}$ or ${\mathbf{10}}$ table~\ref{invariants} shows that the potential has the form \footnote{
Very similar considerations hold in the case ${\mathbf{r_L}}={\mathbf{r_R}}={\mathbf{4}}$, the only change is in the functional form of the leading and
subleading terms.}
\begin{equation}
V^{\mathbf{5+5}}=V_{\rm leading}+V_{\rm sub-leading}=\frac{N_c}{16\pi^2}m_\psi^4\epsilon^2
\left[(a_L+a_R)s_h^2+(b_L\epsilon^2+b_R\epsilon^2)s_h^4\right]\,,
\label{pot5}
\end{equation}
where $a_{L,R}$ and $b_{L,R}$ are model-dependent ${\mathcal{O}}(1)$ numerical coefficients. In the equation above we have assumed, for simplicity,
$\epsilon_L\simeq\epsilon_R=\epsilon=y/g_\psi$, however nothing would be gained if relaxing this assumption. Indeed it is possible to show that the
case $y_L\simeq y_R$ discussed in the present section is the most favorable one, both the fine-tuning and the Higgs mass would increase for large
separation $y_L\ll y_R$ or $y_R\ll y_L$.

The tuning of the Higgs VEV, provided the signs of the coefficients can be freely chosen, requires
\begin{equation}
\left|\frac{a_L+a_R}{b_L\epsilon^2+b_R\epsilon^2}\right|
=2\,\xi\,.
\end{equation}
The amount of cancellation implied by the equation above is
\begin{equation}
{\Delta}^{\mathbf{5+5}}=\frac{{\textrm{max}}(|a_L|,|a_R|)}{|a_L+a_R|}
\simeq\frac1\xi\frac1{\epsilon^2}\,,
\label{tun5}
\end{equation}
and it is parametrically larger than $\Delta_{\rm min}$ for $\epsilon<1$. This accounts for the ``double'' tuning which has to be performed on the potential in eq.~(\ref{pot5}):
one must first cancel the $\epsilon^2$ terms making them of the same order of the formally subleading $\epsilon^4$ ones, and afterwards
further tune the $\epsilon^2$ and $\epsilon^4$ contributions.
Once the minimization condition is imposed we can easily obtain the physical Higgs mass,
\begin{equation}
{m_h^2}\,=\,\frac{8N_cg_\psi^4f^2}{16\pi^2}\xi(1-\xi)\epsilon^4\left(b_L+b_R\right)\simeq \frac{N_c}{2\pi^2}v^2g_\psi^4\epsilon^4\,.
\end{equation}
The advantage of the doubly tuned models, which helps in obtaining a light Higgs boson, is that the Higgs quartic coupling is also automatically
reduced in the tuning process. In spite of the fact that the potential is generated at ${\mathcal{O}}(\epsilon^2)$
indeed the Higgs mass-term scales like $\epsilon^4$ rather than $\epsilon^2$.

\begin{figure}
\centering
\includegraphics[width=.25\textwidth]{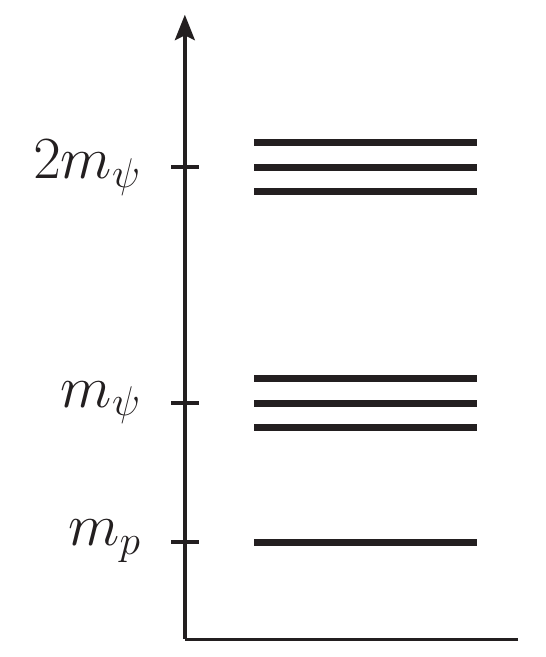}
\caption{Schematic representation of the spectrum of the fermionic resonances.}\label{fig:spectrum}
\end{figure}

However the reduction of $m_h$ is not sufficient for a $125$~GeV Higgs, one extra ingredient is needed. Suppose indeed that we apply the naive estimate of eq.~(\ref{topyukawa})
for the top Yukawa. Since $\epsilon_L\simeq\epsilon_R=\epsilon$ we would obtain $\epsilon\simeq\sqrt{y_t/g_\psi}$ and therefore, taking $g_\psi\sim g_\rho\sim5$ as reference value,
a too
heavy Higgs
\begin{equation}\label{eq:mh_gpsi_5}
m_h^{\mathbf{5+5}}\,\simeq\,\sqrt{\frac{N_c}{2\pi^2}y_t^2 g_\psi^2v^2}=500\,{\textrm{GeV}}\left(\frac{g_\psi}5\right)\,.
\end{equation}
For $g_\psi \gtrsim 2$, a realistic Higgs mass requires that we deviate from
the estimate of eq.~(\ref{topyukawa}), and this can occur if the spectrum of the top
partners is non-generic. Indeed, suppose that one of the partners, with the appropriate
quantum numbers to mix strongly with the left- or right-handed top quark
\footnote{In the cases of the ${\mathbf{5+5}}$ and of the ${\mathbf{10+10}}$ these states
must be in the ${\mathbf{4}}$ and/or in the singlet representation of $\textrm{SO}(4)$.}, becomes
anomalously light, with a mass $m_p$ slightly smaller than $m_\psi$ as depicted in
fig.~\ref{fig:spectrum}.
Given that the Yukawa coupling arises from the mixing with the partners,
its size will be controlled by the mass $m_p$ of the lightest state. Therefore
eq.~(\ref{topyukawa}), that assumes a common mass $m_\psi$ for all the partners,
needs to be modified and becomes~\cite{Matsedonskyi:2012ym}
\begin{equation}
y_t\,\simeq\,y_Ly_R\frac{f}{m_{p}}\,.
\label{ytlp}
\end{equation}
This estimates reduces to eq.~(\ref{topyukawa}) if $m_{p}\simeq m_\psi=g_\psi f$, but it can be parametrically different in the case of a large separation
$m_{p}<m_\psi$. With the above equation and assuming $y_L\simeq y_R$ we obtain $y\simeq \sqrt{y_t m_p/f}$ and therefore a Higgs mass
\begin{equation}
m_h^{\mathbf{5+5}}\,\simeq\,\sqrt{\frac{N_c}{2\pi^2}}\frac{y_tm_p}{f} v=100\,{\textrm{GeV}}\left(\frac{y_tm_p}f\right)\,.
\end{equation}
A realistic Higgs is thus obtained if some of the top partners are light, at least below around $1$ or $2$ times $f$, {\it{i.e.}} $\lesssim2$~TeV for $\xi=0.1$.

No restriction is instead found on the overall scale $m_\psi=g_\psi f$ of the other fermionic resonances. As long as the top partners are light a $125$~GeV
Higgs can be obtained even if $m_\psi$ is large, indeed this mechanism is at work in the 5d holographic models where $m_\psi$ is tied to
$m_\rho\gtrsim3$~TeV. The price to pay, however, is a large tuning. Eq.~(\ref{tun5}) indeed becomes
\begin{equation}
{\Delta}^{\mathbf{5+5}}\simeq \frac1\xi\frac{g_\psi}{y_t}\frac{m_\psi}{m_p}=\frac1\xi\sqrt{\frac{N_c}{2\pi^2}}\frac{g_\psi^2 v}{m_h}\,=\,
\frac1\xi\cdot 20 \left(\frac{125~{\textrm{GeV}}}{m_h}\right)\left(\frac{g_\psi}{5}\right)^2\,,
\label{tundt}
\end{equation}
and the tuning easily overcomes $100$ for a realistic value of $\xi$.

Notice that in the above discussion we have implicitly assumed the existence of a separation among $g_\psi$ and the elementary-composite couplings
$y$, {\it{i.e.}} $y<g_\psi$. The situation is completely different if we instead assume that all the fermionic couplings are of the same order, {\it{i.e.}}
$g_\psi\simeq y\simeq y_t=1$. In this case all the terms in the effective potential become equally large and the issue of double tuning gets nullified, indeed we recover
${\Delta}^{\mathbf{5+5}}=\Delta_{min}$ from eq.~(\ref{tun5}).
Moreover a light Higgs becomes natural (see eq.~(\ref{eq:mh_gpsi_5})) and there is no need to rely on anomalously light partners with specific quantum numbers.
In this case all the fermionic resonances are generically light, with mass $m_\psi\simeq f$, we will consider explicit models with these features in the following section.

\label{dt}

\subsection{Minimal Tuning}

The issue of double tuning appears to be the result of an unfortunate coincidence and not
much effort is needed to avoid it. Indeed it is enough to chose the
fermionic representations in such a way that two or more invariants are allowed in the leading order potential. Sticking to irreducible representations the simplest
choice is ${\mathbf{r_L}}={\mathbf{r_R}}={\mathbf{14}}$. Following table~\ref{invariants} and again assuming $\epsilon_L\simeq\epsilon_R$ the leading
order potential has the form
\begin{equation}
V^{\mathbf{14+14}}=V_{\rm leading}=\frac{N_c}{16\pi^2}m_\psi^4\epsilon^2\left[(a_L+a_R)s_h^2+(b_L+b_R)s_h^4\right]\,,
\label{pot14}
\end{equation}
and it can be adjusted to give a realistic EWSB without need of relying on the subleading terms.
The minimization condition is
\begin{equation}
\left|\frac{a_L+a_R}{b_L+b_R}\right|
=2\,\xi\,,
\end{equation}
and requires a degree of tuning
\begin{equation}
\frac1{\Delta}=\frac{|a_L+a_R|}{{\textrm{max}}(|a_L|,|a_R|)}=2\,\xi\frac{|b_L+b_R|}{{\textrm{max}}(|a_L|,|a_R|)}\,.
\end{equation}
Therefore, in the absence of additional cancellations among $b_L$ and $b_R$, the model has minimal tuning $\Delta^{\mathbf{14+14}}\simeq\Delta_{\rm min}={1}/{\xi}$.

The scenario however becomes problematic when we take into account that the Higgs must be light. Indeed the estimate of $m_h^2$ is now
\begin{equation}
m_h^2\,\simeq \frac{N_c}{2\pi^2}v^2g_\psi^4\epsilon^2\,,
\end{equation}
and it scales like $\epsilon^2$ and not like $\epsilon^4$ as in the case of double tuning. Adopting the naive estimate in eq.~(\ref{topyukawa}) for $y_t$,
which implies $\epsilon\simeq\sqrt{y_t/g_\psi}$, the Higgs is extremely heavy
\begin{equation}
m_h^{\mathbf{14+14}}\,\simeq\,\sqrt{\frac{N_c}{2\pi^2}y_t g_\psi^3v^2}=1\,{\textrm{TeV}}\,\left(\frac{g_\psi}{5}\right)^{3/2}\,.
\label{m1414}
\end{equation}
Of course we could rely on anomalously light top partners to reduce $m_h$ as we did in the case of double tuning. However this mechanism can not reduce $m_h$ indefinitely because the partners can not be arbitrarily light. One unavoidable contribution to their mass comes indeed from the mixing with the elementary states, after diagonalizing the mixing one has $m_p^2=m_*^2+y^2f^2$ where $m_*$ is the contribution to the mass that comes from the strong sector. Even if $m_*$ was taken to vanish we will always have $m_p>yf$ due to the mixing with the elementary states. From eq.~(\ref{ytlp}) we thus obtain
\begin{equation}
y^2\simeq y_t\frac{m_p}{f}\geq y_t y\qquad\Rightarrow\qquad y\geq y_t\,.
\label{boundy}
\end{equation}
Therefore in no case the elementary-composite mixing can go below $y_t$. This bound is not significant in the doubly-tuned case because it corresponds to a
very low Higgs mass. For a minimally tuned model like the one at hand instead the bound gives
\begin{equation}
m_h\,\gtrsim\,\sqrt{\frac{N_c}{2\pi^2}}y_t g_\psi v=500\,{\textrm{GeV}}\,\left(\frac{g_\psi}{5}\right)\,.
\end{equation}
The Higgs is unavoidably too heavy in this class of models even if light top partners are present.

However, notice that the same caution remark given at the end of the previous section applies to the present case as well: when $g_\psi\sim y\sim y_t=1$ all
the issues outlined above disappear. We expect no difficulty in obtaining a light Higgs in this case, the prediction is again that all the fermionic resonances
will have to be light, slightly lighter than the vector ones.

\subsection{Minimal tuning with composite $t_R$}

Another interesting possibility which can alleviate the issue of a too heavy Higgs, is that the $t_R$ is a completely composite chiral state that
emerges from the strong sector. In this case the potential is entirely generated by the left coupling $y_L$. By looking at table~\ref{invariants}
we see that a minimally tuned potential can be obtained also with a completely composite $t_R$ if we assign the left fermionic operator to the
${\mathbf{14}}$. The potential is
\begin{equation}
V^{\mathbf{14}}=V_{\rm leading}=\frac{N_c}{16\pi^2}m_\psi^4\epsilon_L^2\left[a\,s_h^2+b\,s_h^4\right]\,,
\end{equation}
where $a$ and $b$ are, a priori, ${\mathcal{O}}(1)$ coefficients. To tune the electro-weak VEV
we have to require that the coefficient $a$ can be artificially reduced to enforce the condition
\begin{equation}
\left|\frac{a}{b}\right|=2\,\xi\,,
\end{equation}
which corresponds to a cancellation
\begin{equation}
\Delta^{\mathbf{14}}=\frac1{|a|}\simeq\frac{1}{b\,\xi}\,.
\label{tun14tc}
\end{equation}
Provided that no additional cancellation is enforced on $b$ the tuning is minimal
\begin{equation}
\Delta^{\mathbf{14}}\simeq\Delta_{\rm min}=\frac{1}{\xi}\,.
\end{equation}
The Higgs mass-term scales like $\epsilon^2$ as in the previous section, {\it{i.e.}}
\begin{equation}\label{eq:mh14}
m_h^2\,\simeq \frac{N_c}{2\pi^2}bv^2g_\psi^4\epsilon^2\,.
\end{equation}
The difference with the previous case is that now $\epsilon_L$ is smaller, because for a totally composite $t_R$ the top Yukawa
is simply
\begin{equation}\label{eq:mt14}
y_t\simeq y_L\qquad\Rightarrow\qquad \epsilon_L\simeq \frac{y_t}{g_\psi}\,.
\end{equation}
Therefore the Higgs mass is somewhat smaller,
\begin{equation}
m_h^{\mathbf{14}}\,\simeq\,\sqrt{b}\sqrt{\frac{N_c}{2\pi^2}y_t^2 g_\psi^2v^2}=\sqrt{b}\,500\,{\textrm{GeV}}\left(\frac{g_\psi}5\right)\,,
\label{m14tc}
\end{equation}
but not enough.
Notice that no help can come in this case from anomalously light top partners because
the absolute lower bound $y_L\geq y_t$ derived in eq.~(\ref{boundy}) is  already saturated.
We conclude that the Higgs is typically heavy also in the models with total $t_R$ compositeness.
Once again, the only possibility to obtain a $125$~GeV
Higgs with a minimal amount of tuning is to lower the fermionic scale by taking $g_\psi\sim y_L\simeq y_t=1$.

The alternative way to obtain a light Higgs is to reintroduce additional tuning
to lower the Higgs mass. In the case at hand this could be achieved by artificially reducing the parameter
$b$ that controls the Higgs mass (see eq.~(\ref{m14tc})), {\it{i.e.}} by taking
\footnote{An artificial reduction of $m_h$ through the tuning of the quartic term might be
enforced also in the cases --eq.s~(\ref{pot5}),\,(\ref{pot14})-- considered in the previous sections.
We will not discuss this possibility for shortness, and also because it will never be relevant
in the explicit models described in the following. In particular, we find that
in our explicit realization of the $\mathbf{5+5}$ doubly tuned scenario the structure of the
Higgs potential is constrained in a way that an additional cancellation of the quartic cannot
occur at any point of the fundamental parameter space.
}
\begin{equation}
b\simeq \frac1{16}\left(\frac{m_h}{125\,{\textrm{GeV}}}\right)^2\left(\frac5{g_\psi}\right)^2\,.
\end{equation}
This obviously enhances the fine tuning. From eq.~(\ref{tun14tc}) we obtain
\begin{equation}\label{eq:tuning_14}
\Delta \simeq \frac{1}{\xi}\frac{N_c}{2\pi^2}y_t^2 g_\psi^2 \frac{v^2}{m_h^2}
\simeq\frac1{\xi}\cdot 16\left(\frac{125\,{\textrm{GeV}}}{m_h}\right)^2\left(\frac{g_\psi}5\right)^2\,.
\end{equation}

The level of tuning of this scenario is comparable with the one of doubly-tuned models reported in eq.~(\ref{tundt}), however there is a crucial difference
among the two cases. Indeed in the case of section~\ref{dt} the $125$~GeV Higgs requires the existence of anomalously light partners, therefore even
if the fermionic scale $m_\psi$ is high some of the resonances will be light and easily detectable at the LHC. In the present case instead there is no
need of light partners and all the fermionic and gauge resonances could be heavy, lying in the multi-TeV region.
This kind of models evade the  connection among light Higgs and light resonances and they could even escape the direct LHC searches.
Of course they are tuned, but the level  of fine-tuning is comparable with the one of the standard MCHM$_{4,5,10}$ constructions.

\subsubsection*{Double tuning with composite $t_R$}

Another logical possibility that might be considered is the one of doubly tuned models with
composite $t_R$, for example a model where the $q_L$ mixes with a ${\mathbf{5}}$ of
\mbox{SO$(5)$} like the one discussed in Ref.~\cite{marzocca}. Contrary to other models in the literature
in this case tuning of the electo-weak VEV requires a cancellation in the potential between terms
controlled by $y_t^2$ and sub-leading ones proportional to $y_t^4$.  The estimates for this
case are easily extracted from sect.~3.1 by remembering that now $y_L\simeq y_t$, and read
\begin{equation}
{\Delta}^{\mathbf{5}}=\frac1\xi\, \frac{g_\psi^2}{y_t^2}=\frac1\xi\cdot 25\left(
\frac{g_\psi}5
\right)^2\,,\qquad
{m_h}^{\mathbf{5}}=\sqrt{\frac{N_c}{2\pi^2}} y_t^2v\simeq100\,\textrm{GeV}\,.
\end{equation}
Note that in this case the Higgs mass is independent  of the strong sector coupling.
In this setup one thus expects sizable tuning, comparable with the one of the
\mbox{MCHM$_{4,5,10}$}, but no need for anomalously light top partners to
obtain a light enough Higgs. We will not further discuss this option because it
is difficult to realize it an explicit (holographic or deconstructed) calculable model.
In the minimal realizations, indeed, we find that the Higgs potential is too constrained
and that there is not enough freedom in the parameter space to tune $\xi$ to a
realistic value.

\section{Explicit Realizations}

In the previous sections we performed a general model-independent analysis of the fine-tuning in composite Higgs scenarios.
We identified three broad classes of models based on the structure of the Higgs effective potential.
Each class leads to different predictions for the Higgs mass and for the amount of tuning in the Higgs potential,as summarized in table~\ref{tab:stime}.
The aim of the present section is to verify the validity of the general analysis by studying explicit models.
The analysis will also allow us to quantify the amount of deviation one may expect from
the general estimates.

\begin{table}[t]
\renewcommand{\arraystretch}{1.5}
\begin{center}
\begin{tabular}{llll|cccccccc}
\hline
& & & & & $m^2_h$ & & $m_h$\ \ (GeV)& & $\Delta$
\\
\hline
{\rule{0pt}{1.85em}\small Minimal Tuning} & \multicolumn{2}{l}{\hspace{.1cm}\small ${\bf 14}_L +{\bf 14}_R $} && \multicolumn{2}{l}{\hspace{.41cm}\small$\displaystyle\frac {N_c}{2\pi^2} y_t g_\psi^3 v^2$} & \multicolumn{2}{l}{\hspace{.41cm}\small$\displaystyle 125 \left(\frac{g_\psi}{1.2}\right)^{3/2}$} & \multicolumn{2}{c}{\hspace{.31cm}\small$\displaystyle\frac {1}{\xi}$}
\\
{\rule{0pt}{1.85em}\small Double Tuning} & \multicolumn{2}{l}{\hspace{.1cm}\small ${\bf 5}_L + {\bf 5}_R$, ${\bf 10}_L +{\bf 10}_R $}&& \multicolumn{2}{l}{\hspace{.41cm}\small$\displaystyle\frac {N_c}{2 \pi^2} y_t^2 g_\psi^2\, v^2$} & \multicolumn{2}{l}{\hspace{.41cm}\small$\displaystyle 125 \left(\frac{g_\psi}{1.3}\right)$}& \multicolumn{2}{c}{\hspace{.31cm}\small$\displaystyle\frac {g_\psi}{y_t} \frac {1}{\xi}$}
\\
{\rule{0pt}{1.85em}\small Composite $t_R$} & \multicolumn{2}{l}{\hspace{.1cm}\small ${\bf 14}_L +{\bf 1}_R $}&& \multicolumn{2}{l}{\hspace{.41cm}\small$\displaystyle\frac {N_c}{2 \pi^2} y_t^2 g_\psi^2\, v^2$} & \multicolumn{2}{l}{\hspace{.41cm}\small$\displaystyle 125 \left(\frac{g_\psi}{1.3}\right)$} &\multicolumn{2}{c}{\hspace{.31cm}\small$\displaystyle\frac {1}{\xi}$}
\\
{\raisebox{-1.2em}{\rule{0pt}{3.05em}}\small Gauge tuning} & \multicolumn{2}{l}{}&& \multicolumn{2}{l}{\hspace{.41cm}\small$\displaystyle\frac {9}{16\pi^2} g^2 g_\rho^2 v^2 $} & \multicolumn{2}{l}{\hspace{.41cm}\small$\displaystyle 125 \left(\frac{g_\rho}{3.2}\right)$} & \multicolumn{2}{c}{\hspace{.31cm}\small$\displaystyle\frac {9} {8\pi^2} g^2 g_\rho^2 \frac {v^2}{m_h^2} \frac{1}{\xi}$}
\\\hline
\end{tabular}
\end{center}
\caption{\label{tab:stime}\small Estimates for Higgs mass and tuning in various composite Higgs models discussed in section \ref{sec:2a}. For each class the possible embeddings of $q_L$ and $t_R$ are also described. The table only include the minimal tuning required in each scenario and does not take into account possible additional tuning which could lower the Higgs mass.}
\label{summarytable}
\end{table}

We will use the simple but complete 4d implementations of the composite Higgs idea
proposed in Refs.~\cite{Panico:2011pw} and \cite{DeCurtis:2011yx}.
These realizations are  minimal in the sense that only a limited number of composite resonances
are included to ensure the calculability of the Higgs effective potential. Moreover these are
the states potentially accessible at the LHC. Keeping in mind unavoidable differences in the two constructions,
to ensure the finiteness of the effective potential at one-loop level it is sufficient to couple the SM fields
to two $\textrm{SO}(5)$ multiplets, leading to a structure  with ``two levels'' of composite states.
For details on both models see appendix B. In what follows the choice between
the two realizations
is  dictated by convenience. In all cases, we have checked that the two formulations agree within the expected cut-off dependent effects.
The comparison also allows to estimate the model dependence of the results.

In order to evaluate quantitatively the tuning in a given model we adopt the definition of fine-tuning given in Ref.~\cite{bg}
\begin{equation}
\Delta= \max_i\left|\frac  {\partial \log m_Z}{\partial \log x_i}\right|\,,
\label{bgtuning}
\end{equation}
where $x_i$ are the parameters of the theory, and $m_Z= g/\cos(\theta_W) f s_h/2$, which actually establishes the size of $\langle h \rangle$. Keeping fixed $f$ and the gauge couplings, eq.~\eqref{bgtuning} is exactly equivalent to the tuning on $s_h$ and coincides with the definition of tuning usually adopted in the composite Higgs scenarios. The choice of the Barbieri--Giudice measure has been made also in the view of comparing the Composite Higgs tuning with the one of supersymmetric scenarios. For example in the MSSM the tuning is of order 100 or greater, see Ref.~\cite{SUSYtuning} for a partial list of references.

In the CHM  we do not know which are the fundamental
variables of the theory that we should vary to compute the tuning. Nevertheless
we expect that, for a generic choice of the parameters, the logarithmic derivative
will typically reproduce the amount of tuning that we defined in the previous section
as the degree cancellation in the Higgs potential. For the analysis of the explicit models
we will compute the tuning by varying all the parameters of the ``fundamental'' Lagrangian.
For the numerical computation it is useful to notice that
the tuning can be extracted directly from the Higgs potential. Using the minimum condition $V'(s_h)=0$,
the tuning measure can be cast as follows
\begin{equation}\label{tuning}
\Delta= \max_i\left| \frac {2 x_i}{s_h} \frac {c_h^2}{f^2 m_h^2} \frac {\partial^2 V}{\partial x_i \partial s_h}\right|\,.
\end{equation}
Using this formula one can readily derive the tuning estimates of Section 3.

When relevant we will also include in the tuning the gauge contribution.
One interesting point is that the gauge contribution to the potential, often considered sub-leading,
can be relevant in the region of small fermion mass scale, $m_\psi< m_\rho$.
The amount of tuning due to the gauge can be easily estimated.
In the examples that follow the potential can be approximated as
\begin{equation}\label{pot-a1a2}
V\approx \alpha s_h^2 - \beta s_h^2 c_h^2\,.
\end{equation}
The Higgs VEV is determined by the condition $s_h^2=(\beta-\alpha)/(2\beta)$, while the Higgs mass is
\begin{equation}
m_h^2 \simeq \frac{8\beta}{f^4}v^2.
\end{equation}
In the limit $m_\psi< m_\rho$ the gauge loops can give a sizable contribution to
$\alpha$ in eq.~(\ref{pot-a1a2}), and hence to the Higgs mass:
\begin{equation}
\delta m_h \sim \frac{3}{4\pi}g\, g_\rho v = 120 ~{\rm GeV} \left(\frac{g_\rho}{3}\right)\,.
\label{gaugecontribution}
\end{equation}
This contribution is of the size of the measured Higgs mass ($125$ GeV) for $g_\rho\simeq 3$.
Using eq.~(\ref{explicit-pot-gauge}) with $x=m_\rho^2$ we can also quantify
the tuning associated to gauge contributions as
\begin{equation}\label{eq:gaugetuning}
\Delta_{\rm gauge}\approx \frac{1}{\xi}\frac{9}{8\pi^2} g^2 g_\rho^2 \frac {v^2}{m_h^2}\,.
\end{equation}
With obvious identifications of the couplings, one can notice that
the estimate in eq.~(\ref{eq:gaugetuning}) has exactly the same structure of
the fermionic tuning in the minimally tuned
models with composite $t_R$ (see eq.~(\ref{eq:tuning_14})).
Given the bound on the S-parameter, $m_\rho \gtrsim 2.5$~TeV,
eq.~(\ref{eq:gaugetuning}) implies $\Delta \gtrsim 10$ for a realistic Higgs mass.
This is an irreducible source of tuning that exists in all models where the Higgs is a pNGB
even beyond partial compositeness and therefore provides a lower bound.

In what follows we will check the agreement of the numerical results obtained in two
explicit constructions with the general estimates of the tuning and of the Higgs
mass.~\footnote{As an operative definition of the fermionic coupling $g_\psi$ we adopt the quadratic mean of the mass
parameters of the Lagrangians divided by $f$. We have checked that other definitions give comparable results.}
As we will see the agreement is very good in the large-$g_\psi$ region while
some deviation is found in the small-$g_\psi$ region.
We will present our scans for a reference value $f=800$ GeV, corresponding to $\xi=0.1$.
As explained in Ref.~\cite{implications}, these results are easily rescaled to other values of $f$ as long $v/f \lesssim 1/2$.
This can be obtained by rescaling by the same amount all the dimensionful parameters of the Lagrangian.
The fact that the configuration is already tuned, allows us to adjust $\xi = v^2/f^2$ to the desired value by small
perturbations of the parameters. In this way we find points where the Higgs mass remains unchanged,
up to corrections of order $v^2/f^2$. Note also that in so doing the tuning grows proportionally to $f^2$
as it immediately follows from eq.~(\ref{tuning}).

\subsection{Double tuning: CHM$_5$}

As a first explicit example we focus on the class of theories with
``double tuning''. In particular we consider one of the models widely discussed in the
literature, the CHM$_5$, where the SM fermions couple to composite states in the fundamental
$\textrm{SO}(5)$ representation, the $\bf {5}$. The effective Lagrangian for the SM fields
has the general form,
\begin{equation}
\begin{aligned}
{\cal L}_{\rm eff} &=
\bar q_L \slashed{p} \left[ \Pi_0^q
 + \frac{s^2_h}{2} \left( \Pi_1^{q1}\, \widehat H^c \widehat H^{c\dagger}
 +  \Pi_1^{q2}\, \widehat H \widehat H^\dagger \right) \right] q_L \\
& +\bar u_R \slashed{p} \left( \Pi_0^u + \frac{s^2_h}{2}\, \Pi_1^u \right) u_R
 +\bar d_R \slashed{p} \left( \Pi_0^d + \frac{s^2_h}{2}\, \Pi_1^d \right) d_R  \\
&+ \frac{s_hc_h}{\sqrt{2}} M_1^u \,\bar q_L \widehat H^c u_R
 + \frac{s_hc_h}{\sqrt{2}} M_1^d \,\bar q_L \widehat H d_R + h.c. \,.
\end{aligned}
\label{eff5L5R}
\end{equation}
where $\Pi$'s are form factors functions of $p^2$ depending on the model, see appendix.
$\widehat H$ denotes the normalized Higgs doublet: $\widehat H = h_a/(\sum |h_a|^2)^{1/2}$.
The kinetic terms contain a single functional dependence, $s_h^2$. This confirms that the model
belongs to the first class described in section \ref{sec:2a} where subleading terms must
be used to tune the electro-weak VEV.

Let us now compare the numerical results with the general predictions derived
in section~3.
For the scans we chose $\xi = 0.1$ and restricted the composite-fermions
mass parameters to the range $[-10 f, 10 f]$, while the coupling of the gauge resonances
was fixed to the value $g_\rho = 5$. The top mass was set to the value
$m_t = m_t^{\overline {MS}}{(2\ {\rm TeV})} =  150\ {\rm GeV}$,
which corresponds to $m_t^{pole} = 173\ {\rm GeV}$,
and the bottom mass was loosely fixed to the value $m_{b} \simeq 3\ {\rm GeV}$.

\begin{figure}[t!]
\centering
\includegraphics[width=.47\textwidth]{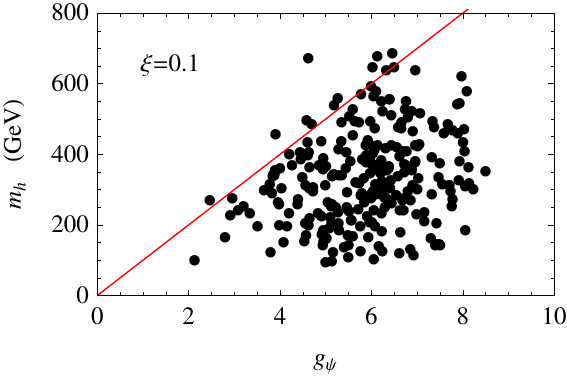}
\hspace{1.5em}
\includegraphics[width=.472\textwidth]{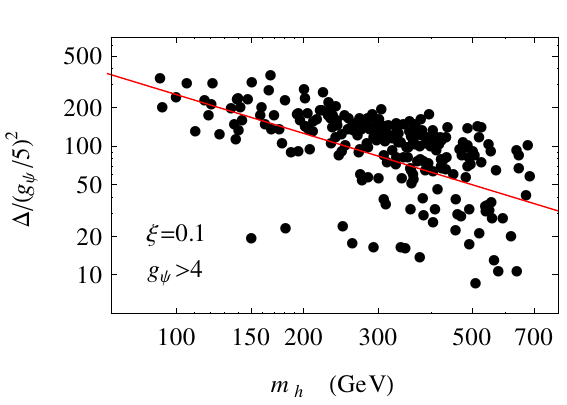}
\caption{\small Scatter plots for the CHM$_5$ set-up corresponding to $\xi=0.1$ for large values of $g_\psi$.
In the left panel we show the Higgs mass as a function of $g_\psi$
and in the right panel the amount of tuning as a function of the Higgs mass.
The red lines correspond the general estimates given in eq.~(\ref{eq:mh_gpsi_5})
and eq.~(\ref{tundt}).
}\label{fig:5+5_large_gpsi}
\end{figure}

First of all we consider the region with large values of $g_\psi$ ($g_\psi \gtrsim 4$).
In this case, in the absence of anomalously light top partners,
the Higgs mass is predicted to be relatively heavy.
The plot in the left panel of fig.~\ref{fig:5+5_large_gpsi} confirms
this expectation and shows that the estimate in eq.~(\ref{eq:mh_gpsi_5})
correctly describes the upper bound of the Higgs mass as a function of $g_\psi$.
Notice that a light Higgs can still be obtained at large $g_\psi$
if some top partners are lighter than the overall scale $m_\psi = g_\psi f$,
as explained in the previous section.~\footnote{An explicit numerical check of this correlation has been presented in Refs.~\cite{Panico:2011pw,Matsedonskyi:2012ym}.}
This comes, however, at the price of a larger tuning, as confirmed by the scatter plot
in the right panel of fig.~\ref{fig:5+5_large_gpsi}.

The parametric estimate of the tuning reported in eq.~(\ref{tundt})
scales quadratically with $g_\psi$. Therefore, to allow a comparison with
the numerical scan, in the scatter plot we normalized the tuning to $(g_\psi/5)^2$.
Since $4 \lesssim g_\psi \lesssim 8$, from the vertical axis we can approximately
read the tuning $\Delta$ of the model. For a light Higgs this is typically above $100$,
as already found to happen in the 5d versions of this model~\cite{Panico:2008bx}.
The estimate is shown as the red line in right plot of fig.~\ref{fig:5+5_large_gpsi}.
We see that it is in fair agreement with
the numerical results which typically fall within a factor $2$ from the estimate.
Some configurations exist, however, in which the logarithmic derivative has a value
significantly below the tuning estimate. This spread is due to peculiar structure of the
leading term of the Higgs effective potential, which can give rise to a sort of ``factorized'' tuning.
The mechanism is simple, the leading term in the effective potential, at least in some regions of the parameter space,
has an approximately factorized structure and the tuning can be achieved by partially cancelling each factor
in an independent way. Although the total amount of cancellation is
always the same at fixed $g_\psi$ and Higgs mass, the logarithmic derivative is not
able to capture this ``factorized'' tuning and gives a smaller value for $\Delta$.

We now consider the region of parameter space with small $g_\psi$, that is with light composite-fermion mass scale. In this case $\epsilon_L \sim \epsilon_R \sim 1$,
thus the expansion in the elementary--composite
mixings breaks down and all the terms appearing in the Higgs effective potential are potentially of the same
order.~\footnote{Notice that, for realistic values of the Higgs compositeness $\xi \ll 1$, the expansion
of the potential as a series in $\sin(h/f)$ is still possible. This implies that
the results of the general analysis of the previous sections remains approximately valid.}
This opens up the possibility to obtain a suitable minimum, with no
additional tuning, by using the sub-leading terms in the potential,
whose size is now comparable with the leading order ones.

Notice that, differently from the large-$g_\psi$ case, in which the top-partners
contributions dominate the Higgs effective potential, in the present situation
the corrections due to the bottom partners and to the gauge fields can have a
significant impact. Indeed the estimates show that, in a typical point of the
parameter space, the contributions of the bottom partners and of the gauge fields
to the Higgs mass can be of order $100\ {\rm GeV}$. In the case of the bottom
this sizable contribution is explained by the fact that configuration with relatively
heavy partners are favored to reduce the corrections $Zb_L \overline b_L$ coupling.
A naive estimate gives
\begin{equation}
\frac{\delta g_{Z b_L \overline b_L}}{g^{SM}_{Z b_L \overline b_L}}
\simeq 2\cdot 10^{-3} \left(\frac{4 \pi}{g^{bottom}_{\psi}}\right) \xi\,,
\end{equation}
suggesting a lower value $g_\psi^{bottom} \gtrsim 5$ on the
scale of the bottom partners in order to satisfy the experimental bounds.
In our numerical analysis we will not impose
a strict bound on the bottom partner masses, but nevertheless we will give a preference
to configuration with a sizable value of $g_\psi^{bottom}$.

\begin{figure}[t!]
\centering
\includegraphics[width=.47\textwidth]{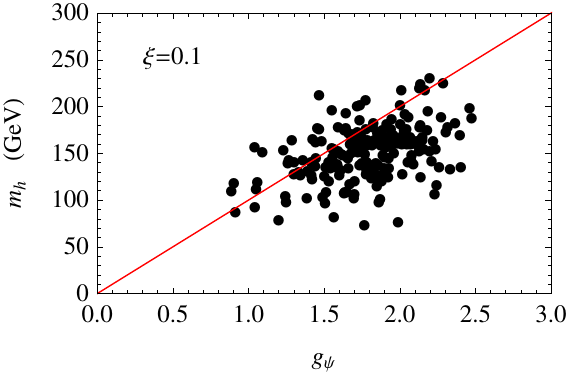}
\hspace{1.5em}
\includegraphics[width=.47\textwidth]{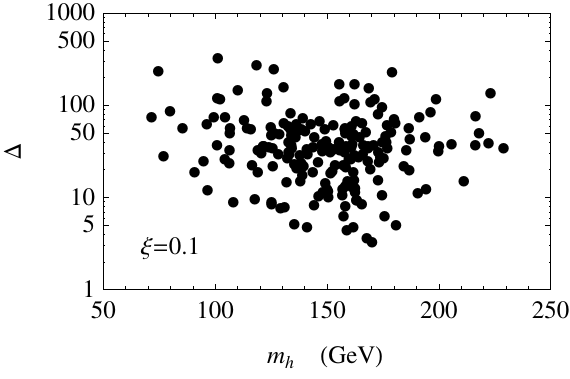}
\caption{\small Scatter plots for the CHM$_5$ set-up corresponding to $\xi=0.1$ for small $g_\psi$
(we restricted the top-partner mass parameters to the range $[-3 f, 3 f]$).
In the left panel we show the Higgs mass as a function of $g_\psi$
and in the right panel the amount of tuning as a function of the Higgs mass for the same
sample points.
The red line in the left panel corresponds to the estimate given in eq.~(\ref{eq:mh_gpsi_5}).
}\label{fig:5+5_small_gpsi}
\end{figure}

As predicted by the estimates (see eq.~(\ref{eq:mh_gpsi_5})), the Higgs mass
can easily take the measured value $m_h \simeq 125\ {\rm GeV}$
if the top partners are light ($g_\psi \lesssim 2$). This can be clearly seen from the scatter
plot in the left panel of fig.~\ref{fig:5+5_small_gpsi}, where the Higgs mass is plotted as a function of $g_\psi^{top}$.
A sizable portion of the parameter space at small $g_\psi$ shows an amount of tuning
in agreement with the estimate $\Delta \sim \xi^{-1}$ (see the plot in the right
panel of fig.~\ref{fig:5+5_small_gpsi}). A significant amount of spread is however present and several
configurations show a degree of tuning much higher than the estimate. Notice that in the plot we only
included the tuning related to the fermionic contribution to the Higgs potential. As discussed previously,
the gauge contribution implies an irreducible tuning $\Delta \gtrsim 10$.

\subsection{Minimal tuning with composite $t_R$}

As a second explicit example, we consider a model belonging to the class of
minimally-tuned scenarios with composite $t_R$.
A set-up with these properties can be realized by coupling the left-handed elementary fermions
to composite states in the symmetric $\textrm{SO}(5)$ representation, the $\bf{14}$.
The new feature of this representation is that its decomposition under $\textrm{SO}(4)$,
\begin{equation}
\mathbf{14}=\mathbf{9}+\mathbf{4}+\mathbf{1}\,,
\end{equation}
contains $3$ representations. As a consequence, $2$ different Higgs dependent structures will appear proportional to the left handed mixing.
The effective action for the SM fields now reads,
\begin{equation}\label{eff14L1R}
\begin{split}
{\cal L}_{\rm eff}&= \bar{q}_L \slashed p \left[\Pi_0^{14_L} + \frac{c_h^2}{2}\Pi_1^{14_L} + \frac{s_h^2}{4}\Pi_1^{14_L} \widehat H^c \widehat H^c + s_h^2 c_h^2 \Pi_2^{14_L} \widehat H^c \widehat H^c \right] q_L  \\
&+\bar{t}_R \slashed p \left[\Pi_0^{1_R} \right] t_R  + s_h c_h \bar{q}_L \widehat H^c\left[ M  \right] t_R + \textrm{h.c.}\,.
\end{split}
\end{equation}
Note that only one field dependence appears in the LR terms,
as also follows from group-theory considerations since only one $\textrm{SO}(4)$ invariant
appears in the product of ${\bf 14}\times {\bf 1}$.
This is an important feature because it avoids dangerous Higgs mediated flavor-changing
neutral currents~\cite{cthdm}.

\begin{figure}[t!]
\centering
\includegraphics[width=.47\textwidth]{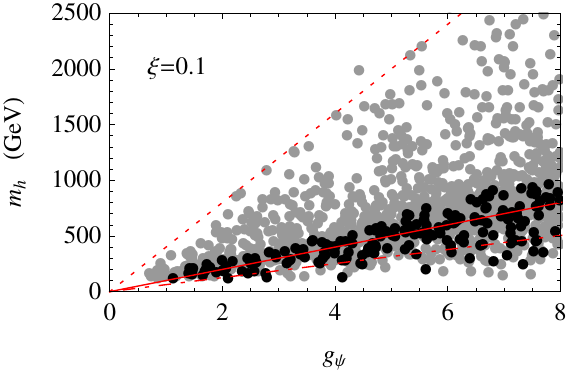}
\hspace{1.5em}
\includegraphics[width=.475\textwidth]{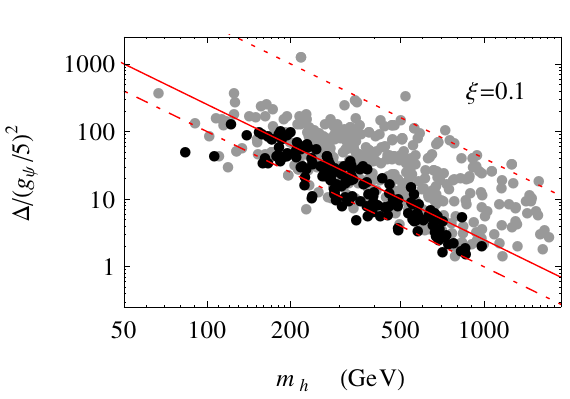}
\caption{\small Left panel: scatter plot of the Higgs mass as a function of $g_\psi$ for $\xi = 0.1$
in the CHM$_{14}$ set-up with composite $t_R$.
Right panel: scatter plot showing the amount of tuning as a function of the
Higgs mass for the same data set.
The solid red lines show the estimates of the Higgs mass and the tuning with $y_L = y_t$,
while the dot-dashed ones correspond to the choice $y_L = \sqrt{2/5}y_t$
and the dotted ones to $y_L = 4 y_t$. The black dots correspond to the points
with $y_L \leq 1$, while the gray ones have $y_L > 1$.
}\label{fig:mh_gstar_14}
\end{figure}

Let us now compare the numerical results with the estimates derived in the
general analysis of the previous sections.
For the scans we set $\xi = 0.1$ and we chose randomly all the composite-sector mass parameters
in the range $[-10 f, 10 f]$ and the elementary--composite mixing $y_L$
in the range $[0, 4 f]$. The top mass if fixed to the value
$m_t = 150\ {\rm GeV}$.

As a first observable we consider the Higgs mass.
The estimates derived in the general analysis predict that a linear correlation
exists between $m_h$ and the fermion mass scale parametrized by $g_\psi$ (see eq.~(\ref{m14tc})).
The numerical analysis, however, shows that a significant amount of spread
is present in the explicit model (see left panel of fig.~(\ref{fig:mh_gstar_14})).
The origin of this deviation can be easily understood.
We verified that the estimate for the Higgs mass in eq.~(\ref{eq:mh14}) is always in good agreement
with the numerical results. On the other hand, the relation between the top mass and the elementary--composite
mixing $y_L$ in eq.~(\ref{eq:mt14}) can be significantly violated.
This can be simply understood by inspecting the approximate expression for the top mass
\begin{equation}
m_t^2 \simeq \frac{5}{16} y_L^2 f^2 \frac{m_R}{m_R^2 + m_{1_{2/3}}^2} \sin^2 \left(\frac{2v}{f}\right)\,,
\end{equation}
where $m_R$ encodes the mass mixing between the $t_R$ and the other composite
states and $m_{1_{2/3}}$ is the mass of the resonance corresponding to the singlet component of the $\bf 14$.
An accidentally small value of the mixing $m_R$ implies a suppression in the top mass, which
must be compensated by a larger value of $y_L$.
Using the approximate analytic expression for the top mass one can derive the bound
\begin{equation}
m_t \lesssim \frac{\sqrt{5}}{2} y_L v\,,
\end{equation}
which implies a lower bound on $y_L$:
\begin{equation}\label{eq:mtyl14}
y_L \gtrsim \sqrt{\frac{2}{5}} y_t \simeq 0.6\,.
\end{equation}
Although the above inequality can be saturated, in a large part of the parameter space
some cancellation occurs and a value of $y_L$ significantly
larger than the minimal one is required. The spread on $y_L$ determines
a corresponding spread in the relation between the Higgs mass and the fermion scale $g_\psi$,
following eq.~(\ref{eq:mh14}).
The estimate is in very good agreement with the numerical results,
as can be seen in fig.~\ref{fig:mh_gstar_14}, where we show the scatter plot
of the Higgs mass as a function of $g_\psi$.
If we restrict our scan to regions with a specific value of $y_L$ then, in the absence of
spurious cancellation, the linear relation given in eq.~(\ref{eq:mh14}) is satisfied.
For instance, in the left panel of fig.~\ref{fig:mh_gstar_14} it is evident that the points
with $y_L \leq y_t$ (the black dots) fall typically in the region predicted by the estimates.

As already pointed out in the general analysis, there
are only two possibilities to get a realistic Higgs mass: considering the
region of the parameter space in which all the fermionic resonances are light
($g_\psi \lesssim 2$), or allow some extra tuning which cancels
the overall size of the effective potential. The relation between the
value of the Higgs mass and the amount of tuning is shown in the right panel
of fig.~\ref{fig:mh_gstar_14},
in which we give the scatter plot of the tuning, defined in eq.~(\ref{bgtuning}), as a
function of the Higgs mass. Also in this case one can see that the spread in the value of $y_L$
determines a corresponding spread in the value of the tuning.
In particular the tuning grows as $y_L^2$ and its estimate can be derived from
the general result in eq.~(\ref{eq:tuning_14}) which corresponds to $y_L \simeq y_t \simeq 1$.
Notice that the configurations in which eq.~(\ref{eq:mtyl14})
is saturated have the smallest possible tuning at fixed Higgs mass.
One can see that the general estimate is well satisfied and an excellent agreement
is obtained if the dependence on $y_L$ is taken into account.

\begin{figure}[t!]
\centering
\includegraphics[width=.48\textwidth]{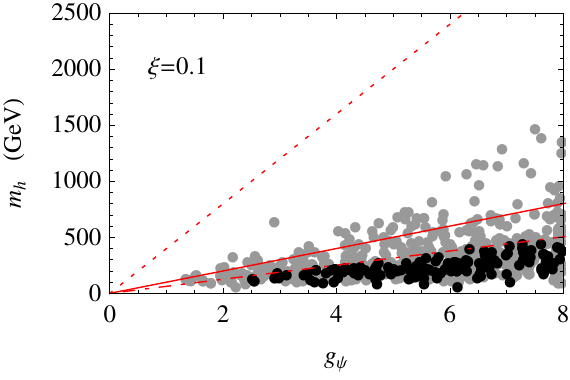}
\hspace{1.5em}
\includegraphics[width=.46\textwidth]{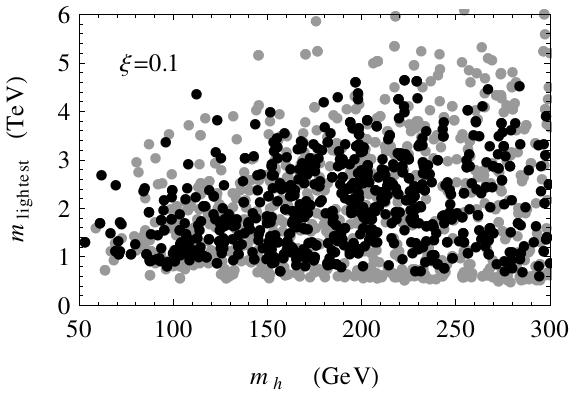}
\caption{\small Scatter plots in the CHM$_{14}$ set-up with composite $t_R$
for $\xi = 0.1$ in the region
of nearly degenerate $\bf 9$ and $\bf 4$ (we allow for a maximal $10\%$ split in the mass parameters). In the left panel we show the Higgs mass as a function of $g_\psi$ and
in the right panel the mass of the lightest fermionic resonance as a function of
the Higgs mass.
For the colors of the points and the meaning of the red lines
see the caption of fig.~\ref{fig:mh_gstar_14}.
}\label{fig:mlight_tuned_14}
\end{figure}

As a final point, we show an example of a region of the parameter space which leads
lo a light Higgs at large $g_\psi$ through some additional tuning. This region is obtained by reducing
the size of the breaking between the $\bf 9$ and the $\bf 4$,
that is by choosing
the mass parameters corresponding to the two representations to be nearly equal.
This choice automatically leads to a cancellation of one of the invariants in the
effective potential and does not imply the presence of light states.
The coefficient of the second invariant must then be tuned to reduce its size and
obtain a suitable minimum. One can see from the scatter plots in
fig.~\ref{fig:mlight_tuned_14} that this region of the parameter
space leads to a realistic Higgs mass
without the necessary presence of light top partners.
The plot on the right shows that, for a realistic Higgs mass, the resonances can be
much heavier than the typical masses required in the double-tuned scenarios,
$m_{lightest} \lesssim 1.5$~TeV (see fig.~\ref{fig:5L5R125} and Ref.~\cite{Matsedonskyi:2012ym}). The plot shows nevertheless a preference for
light states, obviously this is because at fixed Higgs mass points with smaller
$g_\psi$ have lower degree of tuning and are more easily found in the numerical
scan. However the important point is that, at the price of tuning,
resonances as heavy as $4$~TeV can be obtained with a light Higgs.
A model with this feature is very difficult to be discovered.

\section{Implications of a 125 GeV Higgs}

In the light of the recent discovery of a particle compatible with the SM Higgs
and mass around $125$~GeV we now wish to discuss the implications for composite models.
As discussed in the previous sections, a light Higgs can be obtained without
additional tuning only if the composite-fermion mass scale is small
($g_\psi \lesssim 2$). We will focus on this region of the parameter space and
we will present the main features of CHM$_5$ and CHM$_{14}$. As in the previous
analysis our numerical results are obtained for $f=800$ GeV ($\xi = 0.1$), other values of $f$
can be extrapolated as explained in section~4.

As shown in Ref.~\cite{Matsedonskyi:2012ym}, the peculiar structure of the effective
potential in the CHM$_5$ implies the following relation between the Higgs
mass and the mass of the lightest composite fermions which mix directly to the
top~\footnote{The same relation has been obtained by assuming that
the Higgs effective potential satisfies the Weinberg sum rules
in Ref.~\cite{marzocca,pomarol}.}
\begin{equation}
m_h^2 \simeq \frac {N_c}{\pi^2} \frac {m_t^2}{f^2} \frac {m_{2_{1/6}}^2\, m_{1_{2/3}}^2}{m_{2_{1/6}}^2- m_{1_{2/3}}^2}\log \frac {m_{2_{1/6}}^2}{ m_{1_{2/3}}^2}\,,
\label{mh55}
\end{equation}
where $m_{1_{2/3}}$ and $m_{2_{1/6}}$ are the masses of the singlet and of the doublet (including mixing with elementary fields).
It is easy to see why a simple formula holds in this case.
The Higgs mass is determined from the coefficient $\beta$ in eq.~(\ref{pot-a1a2})
that is generated by the top Yukawa contribution to the potential.
In this model only one multiplet of resonances is necessary for the finiteness
of $\beta$ and therefore a formula depending on $m_{1_{2/3}}$ and $m_{2_{1/6}}$ must hold,
at least at leading order in the mixings.
Two multiplets are instead necessary to make $\alpha$ finite.
This however does not affect the Higgs mass due to the fact that $\alpha$ must be
tuned in order to obtain the correct Higgs VEV. Notice that the relation between the
Higgs mass and the lightest resonances masses is a peculiarity of the models with double tuning,
in which one of the invariants has a lower degree of divergence.
In a general case at least two multiplets of resonances are necessary
for finiteness of each term in the potential~\cite{Panico:2011pw,DeCurtis:2011yx}.
As a consequence, there is no guarantee that a simple correlation
of the lightest states and Higgs mass exists. We will see an example below.

\begin{figure}[t!]
\centering
\includegraphics[width=.47\textwidth]{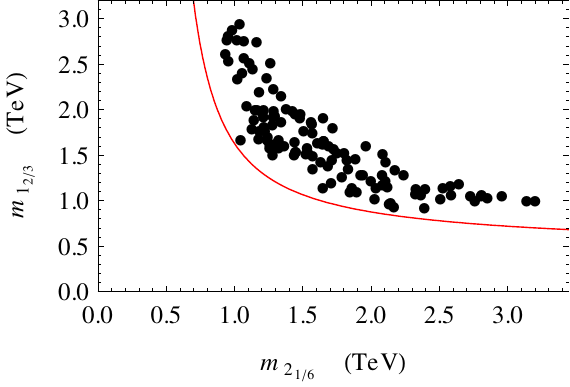}
\hspace{1.5em}
\includegraphics[width=.47\textwidth]{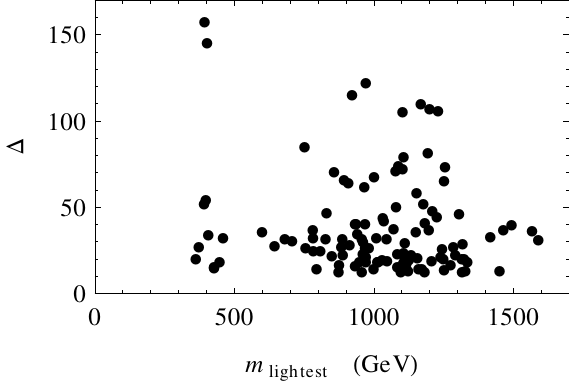}
\caption{\small Scatter plots for $125$~GeV Higgs in CHM$_5$.
We varied the fermionic parameters in the range $0.5 - 5$~TeV and imposed that
the mixings are smaller than $3$. On the left correlation of the
doublet and singlet masses. On the right tuning as a function of the mass
of the lightest resonance.}
\label{fig:5L5R125}
\end{figure}

The correlation in a blind scan between the singlet and the doublet mass is shown on the left plot of fig.~\ref{fig:5L5R125}.
The lightest state is often an exotic doublet with hypercharge $7/6$, the custodian,
that contains an exotic state of electric charge $5/3$. The present experimental bound is
$m_{5/3}\gtrsim 700$ GeV~\cite{Bounds53_CMS,Bounds53_ATLAS} and starts carving out the
natural region of the model. In the right plot the tuning of the various points is considered. We see that
no strong correlation exists between the tuning and the mass of the lightest resonance.
The tuning varies between $10$ and $100$ with an average $\Delta_{avg}=30$.
Note that the lower bound is saturated by the gauge contribution, which
amounts to an irreducible tuning $\Delta \gtrsim 10$.
The tuning is comparable with the tuning of supersymmetric models
with light stops that realize natural SUSY~\cite{SUSYtuning}.~\footnote{To compare with the results
often reported in SUSY literature a factor $2$ in the definition of $\Delta$ should
be taken into account.}

Next let us consider the model with the composite $t_R$ and $q_L$ coupled to fermions
in the {\bf 14}. This was also considered in Ref.~\cite{pomarol} but our results
differ significantly from that analysis. We find that a relation analogous to eq.~(\ref{mh55})
in which the Higgs mass is directly related to the masses of the first level of resonances
does not apply in this case.
The reason for this is the following. The Higgs mass can be determined
by the fermionic contribution to $\beta$. Contrary to CHM$_5$ this is now generated  at leading order
in the mixings and for this reason requires, as $\alpha$, at least two $\textrm{SO}(5)$ multiplets to be
finite.~\footnote{The case discussed in Ref.~\cite{pomarol} is obtained
with a single $\textrm{SO}(5)$ multiplet by tuning parameters of the Lagrangian
to render the contribution to $\beta$ finite. However this does does not hold
in a generic point of the parameter space in our construction.}
As a consequence the potential is sensitive to the second layer of resonances
and no simple correlation will hold among the lighter states.

\begin{figure}[t!]
\centering
\includegraphics[width=.48\textwidth]{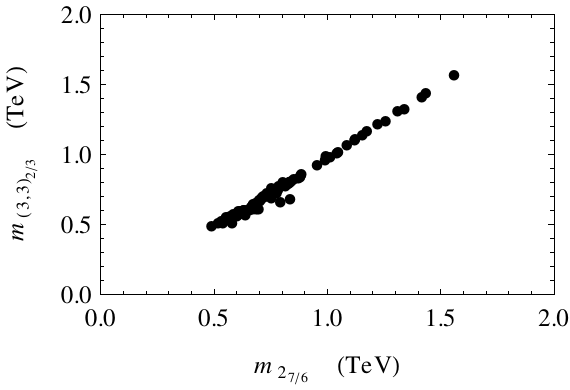}
\hspace{1.5em}
\includegraphics[width=.455\textwidth]{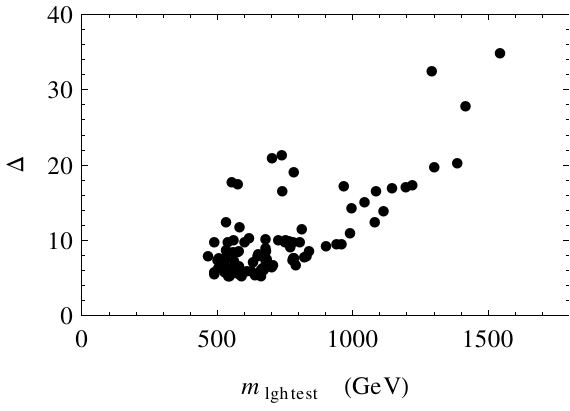}
\caption{\small Scatter plot for 125 GeV Higgs in CHM$_{14}$. The gauge contribution is
computed with $m_\rho=m_{a_1}/\sqrt{2}=2.5$~TeV.
On the left correlation between the $2_{7/6}$ and $(3,3)_{2/3}$ states. On the right fermionic contribution to the tuning as a function of the lightest fermionic resonance.}
\label{fig:14L1R125}
\end{figure}

In fig.~\ref{fig:14L1R125} we consider a particular region of parameter space corresponding to the coupling of spin-1 resonances $g_\rho\approx3$.
We find in this case that $m_{2_{7/6}}$ and  $m_{(3,3)_{2/3}}$ are almost degenerate. The result can be simply understood.
The gauge loops contribute to the coefficient $\alpha$ in the potential and from eq. (\ref{gaugecontribution}) this in isolation (upon tuning the electro-weak VEV)
produces a light Higgs for $g_\rho\approx3$. In CHM$_{14}$ for $m_{2_{7/6}} = m_{(3,3)_{2/3}}$ the fermionic contribution to $\alpha$ exactly vanishes due to
an enhanced symmetry of the fermionic sector in this limit ($\textrm{SU}(13)$ explicitly broken by gauge interactions).
The small correction necessary to obtain a 125 GeV Higgs then requires  $m_{2_{7/6}} \approx m_{(3,3)_{2/3}}$.

Concerning the tuning, the fermionic contribution is shown in fig.~\ref{fig:14L1R125} on the right.
It is typically smaller than in CHM$_5$. This model realizes therefore the minimal tuning, given by the gauge contribution.
However it should be kept in mind that this conclusion relies on our definition of tuning that identifies as one of the variables
the splitting between $m_{2_{7/6}}$ and $m_{(3,3)_{2/3}}$, see our basis of operators in Appendix~A.2.
As a consequence $m_{2_{7/6}}\sim m_{(3,3)_{2/3}}$ does not worsen the tuning. We can attach the following physical meaning
to this: when $m_{2_{7/6}}=m_{(3,3)_{2/3}}$ the fermionic sector of the theory acquires an enhanced global symmetry.
With a different choice of operators or different parameters in the gauge sector
we expect a similar tuning as the in the CHM$_5$.~\footnote{Notice that in the configurations with enhanced global symmetry
only one of the two invariants in the Higgs potential vanishes, while the size of the other still respects the general
estimates. As a consequence, for a sizable value of $g_\psi$, a large tuning is still necessary in agreement with the general results.
The presence of the enhanced symmetry can only mildly improve the amount of tuning, but not eliminate it. In the plots shown
in this section the small amount of tuning is a consequence of the small values for the fermionic scale we used in the scans ($g_\psi \lesssim 2$).}

\section{Conclusions}

We investigated quantitatively the tuning of composite Higgs models with partial compositeness and the interplay
with the predicted value of the Higgs mass. The tuning is often estimated as $1/\xi=f^2/v^2$. While this is the universal scaling in reality the situation
is more complex and depends on the quantum numbers of the composite fermions to which the SM fermions couple.
We identified three classes of models, summarized in table~\ref{summarytable},
characterized by the type of cancellation required to generate the electro-weak VEV.
Within each class the expected size of the Higgs mass can be different and thus the recent
discovery of a light Higgs can have a different impact. For the models in the second and
third class it is difficult to obtain a light enough Higgs for a large strong  sector coupling.

However the tension with the observed Higgs mass disappears in the limit
of light fermionic scale,
corresponding to $g_\psi=m_\psi/f\sim1$. In this case a light Higgs is easily obtained and
also the double tuning issue encountered for the models in the first category tends to
disappear. In the limit of small $g_\psi$ all the models become equivalent for what the
structure of the Higgs potential is concerned and the three classes basically merge in a single
one. When $g_\psi$ is weak many options open up to build models with a light Higgs and
a mild tuning. In section~4 and 5 we studied two examples of such a model: the one based
on the ${\bf{5}}_L+{\bf{5}}_R$ representation and the one with a ${\bf{14}}_L$ and
completely composite $t_R$.

With our classification we found that the only way to obtain a light enough Higgs
with moderate tuning is to work at low $g_\psi$, {\it{i.e.}} to assume a low scale for the
fermionic resonances.
The implication is that light fermionic colored resonances, the top partners, are an expected feature of the
composite Higgs models. Not observing these particles at the LHC would rapidly
carry the scenario in a finely-tuned territory.
In this respect our set-up is similar to the Natural SUSY construction, in which
one requires the stop to be lighter than the other supersymmetric states~\cite{Cohen:1996vb}.
Indeed the
amount of tuning is comparable in the two cases. However in Natural SUSY one relies
on additional non-minimal contributions to obtain a heavy enough Higgs. In our
case instead the model remains minimal and no other contributions to the Higgs potential
are required besides the ones from the top and the gauge sectors.
The need of light states for a moderate tuning is one further motivation for a
serious program of experimental top-partners search at the LHC. At present the stronger bound is
on the charge $5/3$ state which is part of the bi-doublet \cite{Bounds53_CMS,Bounds53_ATLAS}.
A study of the available constraints will be presented by one of us in Ref.~\cite{AAAR}.

\begin{figure}[t!]
\centering
\includegraphics[width=.7\textwidth]{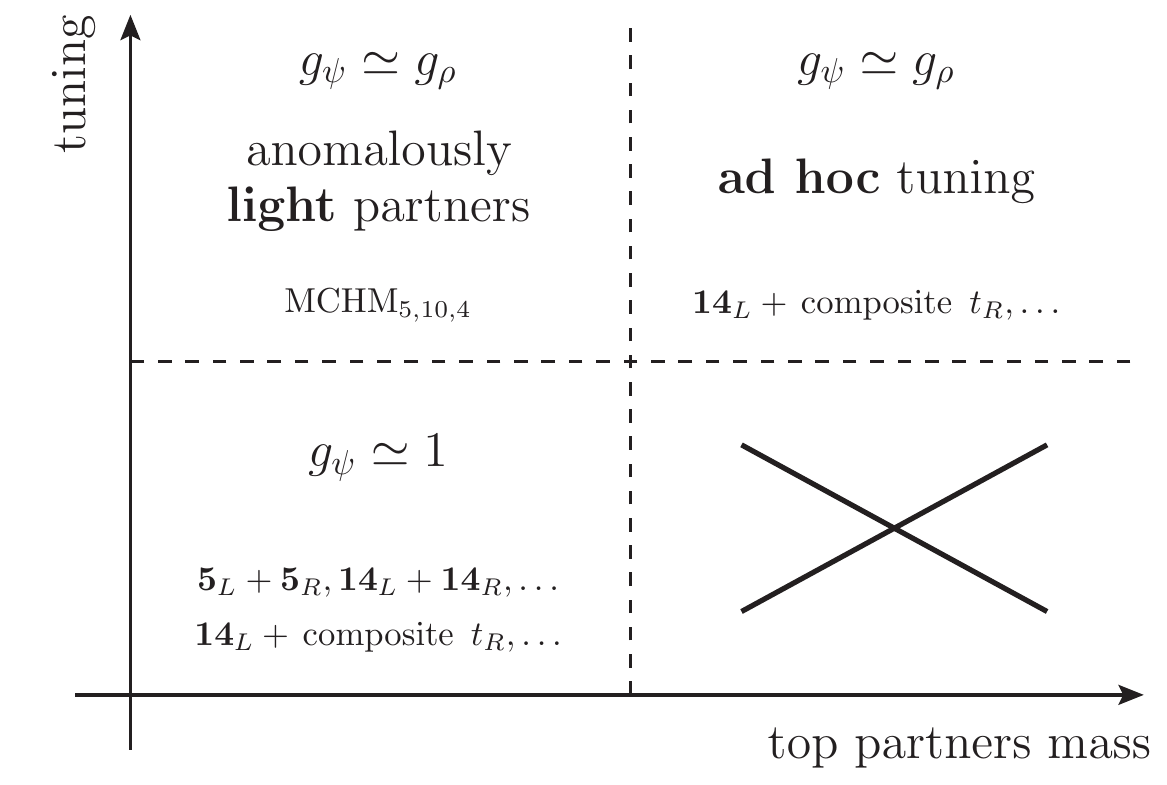}
\caption{\small Schematic representation of the properties of the three basic classes
of composite-Higgs models.}
\label{fig:Tuning_table}
\end{figure}

We also considered the possibility of a larger tuning (100 or larger).
In this case we found two possibilities to obtain a realistic Higgs mass, as summarized
in fig.~\ref{fig:Tuning_table}. One option is to stick to models with doubly-tuned potential like the standard
MCHM$_{4,5,10}$. In this scenario a light Higgs requires the presence of light top partners significantly
below the typical fermionic-resonance scale. The spectrum is characterized by one or two light
multiplets, a fourplet or a singlet of SO$(4)$, while all the other resonances are heavy and
lie in the mass range of the vectors, $m_\rho>2.5$~TeV. In the case of low $g_\psi$
previously discussed, instead, all the fermionic resonances are light and they
can have different quantum numbers. For instance in the model with the ${\bf{14}}$
we expect light top partners in the ${\bf{4}}$, in the ${\bf{1}}$  and in the ${\bf{9}}$ of
SO$(4)$. Alternatively, for similar tuning, one can also have models with heavy fermionic
resonances where the Higgs mass is tuned independently of the electro-weak VEV, we provided one example based on the ${\bf{14}}_L$ and totally
composite $t_R$. This case is indicated in the upper right corner of fig.~\ref{fig:Tuning_table}.
A model of this kind is rather difficult to test directly at the LHC, therefore if no top
partners are found it might become the last corner where the Composite Higgs
scenario could hide.

Our results also have theoretical implications. If we insist on a moderate tuning we need a
separation among the mass scale of the fermonic and of the gauge resonances, and it is
not easy to imagine the origin of this separation. For examples in the models with a
geometrical origin, like the 5d holographic ones, the mass of the fermions is typically
tied to the one of the vectors since both originate from the compactification length of the
space. Therefore it is difficult to describe the separation with 5d models, indeed in this
paper we employed non-geometrical 4d constructions where the fermonic and gauge masses
are independent parameters. However at the fundamental level the problem remains.
``Normal'' strongly-coupled theories like QCD are characterized by a unique scale of
confinement and all the resonances (aside from the baryons in the large-$N_c$ limit)
are expected to have comparable masses. Moreover to obtain a light Higgs we are led
to consider rather low masses, that correspond to a weak fermonic coupling
$g_\psi\sim1$. Thus the interpretation of our models in terms of a strongly-coupled
dynamics could be improper. It would be interesting to identify a possible
UV-completion of these constructions.

\subsection*{Acknowledgments}

We thank R.~Rattazzi for useful discussions.
A. W. was partly supported by the ERC Advanced Grant No. 267985 Electroweak Symmetry
Breaking, Flavour and Dark Matter: One Solution for Three Mysteries (DaMeSyFla)
and by the European Programme Unication in the LHC Era, contract PITN-GA-2009-237920 (UNILHC).

\appendix

\section{Structure of the explicit models}
\label{appA}

The general structure of calculable composite Higgs models was introduced in refs.~\cite{Panico:2011pw,DeCurtis:2011yx} and
we refer to these references for all details. The minimum number of states required to achieve a finite potential at 1-loop is two complete $\textrm{SO}(5)$
multiplets.
Explicit formulas can be found in the references above for the CHM$_5$.

We present in this appendix the relevant models with the $t_L$ coupled
to fermions in the {\bf 14} and a totally composite $t_R$. We introduce two Dirac fermions
in the ${\bf 14}$ representation, $\psi$ and $\widetilde{\psi}$. A suitable basis is given by
symmetric traceless $5\times 5$ matrices. Under the  $\textrm{SO}(4)$ subgroup they decompose as
\begin{equation}
\begin{split}
\mathbf{(3,3)}: \qquad&
\left\{
\begin{array}{ll}
T^{aa}_{ij} = \displaystyle \delta^a_i \delta^a_j + \delta^4_i \delta^4_j
- \frac{1}{2} \delta^i_j\,, \qquad& a=1,2,3\\
\rule{0pt}{1.75em}T^{ab}_{ij} = \displaystyle\frac{1}{\sqrt{2}}
(\delta^a_i\delta^b_j+\delta^a_j\delta^b_i)\,,\qquad&  a<b\,, \quad a,b=1,..,4
\end{array}
\right.\\
\mathbf{(2,2)}: \qquad& \hat T^a_{ij}=\frac{1}{\sqrt{2}}(\delta^a_i\delta^5_j+\delta^a_j\delta^5_i)\,,\qquad\quad\ \ \, a=1,..,4\\
\mathbf{(1,1)}: \qquad& \hat T^0= \frac{1}{2\sqrt{5}}\mathrm{diag}(1,1,1,1,-4)\,.
\end{split}
\end{equation}

\subsection{Discrete Composite Higgs Model}

The schematic structure of the theory can be visualized as a three-site
model (see Ref.~\cite{Panico:2011pw}).
The underlying symmetry is given by two non-linear $\sigma$-models based on the
symmetry breaking pattern
$\textrm{SO}(5)_L \times \textrm{SO}(5)_R/\textrm{SO}(5)_D$.~\footnote{An extra $U(1)_X$
global symmetry must be included to accommodate the fermion hypercharges.
See Refs.~\cite{Panico:2011pw,DeCurtis:2011yx} for further details.}
This structure gives rise to two sets of Goldstone bosons $U_{1,2}$, which are
in part reabsorbed by the gauge fields which gauge some subsets of the global
symmetries at each site. The net result is a theory containing only the $4$
Goldstone bosons corresponding to the usual Higgs doublet.

Each fermionic state is associated to one of the sites.
In particular the first site is associated to the elementary fields,
while the other two are related to the composite states.
The vector-like composite resonances at the middle and last site
$\psi$ and $\widetilde \psi$ are embedded in the representation ${\bf 14}$.
In addition to the vector-like resonances, the composite $t_R$ is embedded
in a total singlet at the last site. At the last site we also
allow for a breaking of the $\textrm{SO}(5)_R$ global symmetry,
in such a way to preserve only an $\textrm{SO}(4)$ subgroup.
This explicit breaking, obtained through mass terms,
encodes the $\textrm{SO}(5) \rightarrow \textrm{SO}(4)$
spontaneous symmetry breaking of the strong sector.

The Lagrangian for the composite
states in the ``holographic gauge''~\footnote{In this gauge the Goldstone
matrix $U_2$ is set to the identity and the physical Goldstones are encoded in
the matrix $U = U_1$.
The terminology ``holographic gauge''
is derived from the holographic technique in extra-dimensional theories~\cite{Panico:2007qd}.
For a discussion on how to reach this gauge see Ref.~\cite{Panico:2011pw}.} reads
\begin{eqnarray}
{\cal L}^{\textrm f}_{\textrm{comp}} &=& \textrm{Tr}[\,i\, \overline \psi \Dslash \psi - m \overline \psi \psi]\nonumber\\
&& +\, \textrm{Tr}[\,i\, \overline {\widetilde\psi} \Dslash \widetilde\psi
- \widetilde m_\Phi \overline {\widetilde\Phi} \widetilde\Phi
- \widetilde m_Q \overline {\widetilde Q} \widetilde Q
- \widetilde m_T \overline {\widetilde T} \widetilde T]\nonumber\\
&& +\, \textrm{Tr}[\,i\, \overline t_R \Dslash t_R - m_R \overline t_R \widetilde T_L] + \textrm{h.c.}\nonumber\\
&& -\, \Delta \textrm{Tr}[\overline \psi \widetilde \psi] + \textrm{h.c.}\,,
\label{eq:compLagr14}
\end{eqnarray}
where we denoted by $\widetilde \Phi$, $\widetilde Q$ and $\widetilde T$
respectively the $(\bf{3}, \bf{3})$, $(\bf{2}, \bf{2})$ and $(\bf{1}, \bf{1})$
components of the $\widetilde \psi$ multiplet.
The mixing in the last line of the above equation comes from a mixing term
involving the $U_2$ Goldstone matrix,
$\Delta \textrm{Tr}[U_2^\dagger \overline \psi U_2 \widetilde \psi] + \textrm{h.c.}$,
which appears in the non-gauge-fixed Lagrangian.

The fermions at the first site only include the $q_L$ elementary doublet.
The Lagrangian for the elementary states is
\begin{equation}
{\cal L}^{\textrm f}_{\textrm{elem}} = i\, \overline q_L \Dslash q_L
- y_L f_\pi \textrm{Tr}[U^\dagger \overline q^{\bf 14}_L U \psi_R] + \textrm{h.c.}\,,
\end{equation}
where $q^{\bf 14}_L$ denotes the embedding of the elementary doublet $q_L$ into the bidoublet
of the ${\bf 14}$ representation.
The Goldstone decay constant $f_\pi$ is related to the decay constants of the two
original non-linear $\sigma$-models by $f_\pi = f/\sqrt{2}$.

\subsection{Minimal 4D Composite Higgs}

The Lagrangian with  composite $t_R$ can be obtained by a slight modification of
the setup of Ref.~\cite{DeCurtis:2011yx} where the $t_R$ is coupled with the strong sector fields
in an $\textrm{SO}(5)$ invariant fashion. In this section we briefly review the model discussed in that paper.

The setup is a two-site model: a $\sigma$ model $\textrm{SO}(5)_L\times \textrm{SO}(5)_R/\textrm{SO}(5)_{L+R}$
parametrized by the unitary matrix $\Omega$ and a second one $\textrm{SO}(5)_2/\textrm{SO}(4)$ parametrized by the vector $\Phi$.
Resonances are introduced as $\textrm{SO}(5)$ gauge fields by gauging the diagonal subgroup of $\textrm{SO}(5)_{R}$ and $\textrm{SO}(5)_2$.  In the unitary gauge, before gauging the SM symmetry, we have a massless fourplet of scalar fields with quantum numbers of the Higgs doublet, while the orthogonal combination of GB's forms the longitudinal components of the $\rho$'s.
In the fermion sector, each SM fermion chirality is coupled to a complete Dirac $\textrm{SO}(5)$ multiplet, which can occur in any representation. Only couplings with certain chirality are retained similarly to 5D models.

For CHM$_{14}$ with composite $t_R$ the action reads,
\begin{equation}
\label{14L14R}
\begin{aligned}
{\cal L}_{\mathbf{14}_L + \mathbf{1}_R} &=\bar{q}^{\rm el}_L i \slashed{D}^{\rm el} q^{\rm el}_L \\
&+ \Delta_{q_L} {\rm Tr}\left[\Omega^\dagger \bar{q}^{\rm el}_L \Omega \psi_T \right] + \textrm{h.c.}\\
&+ {\rm Tr}\left[\bar{\psi}_T \left(i\slashed{D}^{\rho} - m_T \right)\psi_T \right] + {\rm Tr}\left[\bar{\psi}_{\widetilde T} \left(i\slashed{D}^{\rho} - m_{\widetilde T}  \right)\psi_{\widetilde T} \right] \\
&+ \bar{t}^{\rm comp}_R i \slashed{D}^{\rho} t^{\rm comp}_R  + \Delta_{t_R} \Phi^T \bar\psi_{T} \Phi t^{\rm comp}_R + \textrm{h.c.}\\
&- Y_1 \Phi \bar{\psi}_{T,L} \psi_{\widetilde T, R}\Phi  - Y_2 \Phi^T \bar{\psi}_{T,L}\Phi \Phi^T\psi_{\widetilde T, R}\Phi- Y_3{\rm Tr}\left[ \bar{\psi}_{T,L} \psi_{\widetilde T, R} \right]+ \textrm{h.c.}\,.
\end{aligned}
\end{equation}
where $\psi_{T,\widetilde{T}}$ are Dirac fermions in the $\mathbf{14}$. For simplicity we ignore a possible coupling of $t_R$ with $\psi_{\widetilde{T}}$.
Integrating out the strong sector fields one obtains the following expressions for the form factors appearing in \eqref{eff14L1R},
\begin{equation}
\label{matching}
\begin{split}
\Pi_0^{14_L}&=\Pi_{LL}[m_T, m_{\widetilde{T}},Y_3],\quad\quad \Pi_1^{14_L}=2 (\Pi_{LL}[m_T, m_{\widetilde{T}},Y_1/2+Y_3] -\Pi_{LL}[m_T, m_{\widetilde{T}},Y_3]) \\
\Pi_2^{14_L}&=\frac{5}{4}\Pi_{LL}[m_T, m_{\widetilde{T}},\frac{4(Y_1+Y_2)}{5}+Y_3]-2 \Pi_{LL}[m_T, m_{\widetilde{T}},Y_1/2+Y_3]+\frac{3}{4}\Pi_{LL}[m_T, m_{\widetilde{T}},Y_3] \\
\Pi_0^{1_R}&=\Pi_{RR}[m_T, m_{\widetilde{T}},\Delta],\quad\quad M=M[m_T, m_{\widetilde{T}},\Delta]
\end{split}
\end{equation}
where
\begin{equation}
\begin{split}
{\Pi}_{LL}[m_1,m_2, m_3]&= \frac{\left(m_2^2+m_3^2-p^2\right)\,\Delta^2}{p^4 - p^2 (m_1^2+m_2^2+m_3^2) +m_1^2m_2^2}\\
{\Pi}_{RR}[m_1,m_2, m_3]&= \frac{\left(m_2^2-p^2\right)\,\Delta^2}{p^4 - p^2 (m_1^2+m_2^2+m_3^2) +m_1^2m_2^2}\\
{M}[m_1,m_2, m_3]&=\frac {m_1(m_2^2-p^2)}{p^4-p^2(m_1^2+m_2^2+m_3^2)+m_1^2 m_2^2}
\end{split}\,.
\end{equation}
From here the Higgs potential can be computed as explained in Ref.~\cite{DeCurtis:2011yx}.

If one does not introduce $\psi_{\widetilde{T}}$, the action has an accidental symmetry due to which $\Pi_1^{14_L}$ vanishes. If this is the case the electro-weak
VEV must be tuned with the gauge contribution. A light Higgs mass is obtained for $g_\rho\sim 3$.
To be general in our plots we consider the model described in eq.~\eqref{14L14R}.


\begin{thebibliography}{99}

\bibitem{:2012gk}
  G.~Aad {\it et al.}  [ATLAS Collaboration],
  ``Observation of a new particle in the search for the Standard Model Higgs boson with the ATLAS detector at the LHC,''
  Phys.\ Lett.\ B {\bf 716} (2012) 1
  [arXiv:1207.7214 [hep-ex]].

\bibitem{:2012gu}
  S.~Chatrchyan {\it et al.}  [CMS Collaboration],
  ``Observation of a new boson at a mass of 125 GeV with the CMS experiment at the LHC,''
  Phys.\ Lett.\ B {\bf 716} (2012) 30
  [arXiv:1207.7235 [hep-ex]].

\bibitem{pc}
 D.~B.~Kaplan,
  ``Flavor at SSC energies: A New mechanism for dynamically generated fermion masses,''
  Nucl.\ Phys.\ B {\bf 365} (1991) 259.
R.~Contino, T.~Kramer, M.~Son and R.~Sundrum,
  ``Warped/composite phenomenology simplified,''
  JHEP {\bf 0705} (2007) 074
  [hep-ph/0612180].

\bibitem{gk}
  D.~B.~Kaplan, H.~Georgi,
  ``SU(2) x U(1) Breaking by Vacuum Misalignment,''
  Phys.\ Lett.\  {\bf B136 } (1984)  183.
  D.~B.~Kaplan, H.~Georgi, S.~Dimopoulos,
  ``Composite Higgs Scalars,''
  Phys.\ Lett.\  {\bf B136 } (1984)  187.
  H.~Georgi, D.~B.~Kaplan, P.~Galison,
  ``Calculation Of The Composite Higgs Mass,''
  Phys.\ Lett.\  {\bf B143 } (1984)  152.
  T.~Banks,
  ``Constraints on SU(2) x U(1) breaking by vacuum misalignment,''
  Nucl.\ Phys.\  {\bf B243 } (1984)  125.
  H.~Georgi, D.~B.~Kaplan,
  ``Composite Higgs and Custodial SU(2),''
  Phys.\ Lett.\  {\bf B145 } (1984)  216.
  M.~J.~Dugan, H.~Georgi, D.~B.~Kaplan,
  ``Anatomy of a Composite Higgs Model,''
  Nucl.\ Phys.\  {\bf B254 } (1985)  299.

\bibitem{Agashe:2004rs}
  K.~Agashe, R.~Contino and A.~Pomarol,
  ``The Minimal composite Higgs model,''
  Nucl.\ Phys.\ B {\bf 719} (2005) 165
  [hep-ph/0412089].

\bibitem{Contino:2006qr}
  R.~Contino, L.~Da Rold and A.~Pomarol,
  ``Light custodians in natural composite Higgs models,''
  Phys.\ Rev.\ D {\bf 75} (2007) 055014
    [hep-ph/0612048].

\bibitem{DP}
S.~Dimopoulos, J.~Preskill,
  ``Massless Composites With Massive Constituents,''
  Nucl.\ Phys.\  {\bf B199 } (1982)  206.

\bibitem{Panico:2006em}
  G.~Panico, M.~Serone and A.~Wulzer,
  ``Electroweak Symmetry Breaking and Precision Tests with a Fifth Dimension,''
  Nucl.\ Phys.\ B {\bf 762} (2007) 189
  [hep-ph/0605292].

\bibitem{Panico:2008bx}
  G.~Panico, E.~Ponton, J.~Santiago and M.~Serone,
  ``Dark Matter and Electroweak Symmetry Breaking in Models with Warped Extra Dimensions,''
  Phys.\ Rev.\ D {\bf 77} (2008) 115012
  [arXiv:0801.1645 [hep-ph]].

\bibitem{Csaki:2008zd}
  C.~Csaki, A.~Falkowski and A.~Weiler,
  ``The Flavor of the Composite Pseudo-Goldstone Higgs,''
  JHEP {\bf 0809} (2008) 008
  [arXiv:0804.1954 [hep-ph]].

\bibitem{cthdm}
  J.~Mrazek, A.~Pomarol, R.~Rattazzi, M.~Redi, J.~Serra and A.~Wulzer,
  ``The Other Natural Two Higgs Doublet Model,''
  Nucl.\ Phys.\ B {\bf 853} (2011) 1
  [arXiv:1105.5403 [hep-ph]].

\bibitem{Matsedonskyi:2012ym}
  O.~Matsedonskyi, G.~Panico and A.~Wulzer,
  ``Light Top Partners for a Light Composite Higgs,''
  arXiv:1204.6333 [hep-ph].

\bibitem{Barbieri:2000gf}
  R.~Barbieri and A.~Strumia,
  ``The 'LEP paradox',''
  hep-ph/0007265.

 \bibitem{SILH}
  G.~F.~Giudice, C.~Grojean, A.~Pomarol and R.~Rattazzi,
  ``The Strongly-Interacting Light Higgs,''
  JHEP {\bf 0706} (2007) 045
  [hep-ph/0703164].

\bibitem{Panico:2011pw}
  G.~Panico and A.~Wulzer,
  ``The Discrete Composite Higgs Model,''
  JHEP {\bf 1109} (2011) 135
  [arXiv:1106.2719 [hep-ph]].

\bibitem{DeCurtis:2011yx}
  S.~De Curtis, M.~Redi and A.~Tesi,
  ``The 4D Composite Higgs,''
  JHEP {\bf 1204} (2012) 042
  [arXiv:1110.1613 [hep-ph]].

\bibitem{bg}
  R.~Barbieri and G.~F.~Giudice,
  ``Upper Bounds on Supersymmetric Particle Masses,''
  Nucl.\ Phys.\ B {\bf 306} (1988) 63.

 \bibitem{SUSYtuning}
  M.~W.~Cahill-Rowley, J.~L.~Hewett, A.~Ismail and T.~G.~Rizzo,
  ``The Higgs Sector and Fine-Tuning in the pMSSM,''
  arXiv:1206.5800 [hep-ph];
H.~Baer, V.~Barger, P.~Huang, D.~Mickelson, A.~Mustafayev and X.~Tata,
``Post-LHC7 fine-tuning in the mSUGRA/CMSSM model with a 125 GeV Higgs boson,''  arXiv:1210.3019 [hep-ph];
C.~Wymant, ``Optimising Stop Naturalness,''  arXiv:1208.1737 [hep-ph];
S.~Cassel and D.~M.~Ghilencea,  ``A Review of naturalness and dark matter prediction for the Higgs mass in MSSM and beyond,''
  Mod.\ Phys.\ Lett.\ A {\bf 27}, 1230003 (2012)
  [arXiv:1103.4793 [hep-ph]].
 L.~J.~Hall, D.~Pinner and J.~T.~Ruderman,
  ``A Natural SUSY Higgs Near 126 GeV,''
  JHEP {\bf 1204}, 131 (2012)
  [arXiv:1112.2703 [hep-ph]].
  A.~Arvanitaki and G.~Villadoro,
  ``A Non Standard Model Higgs at the LHC as a Sign of Naturalness,''
  JHEP {\bf 1202}, 144 (2012)
  [arXiv:1112.4835 [hep-ph]].

\bibitem{implications}
  M.~Redi and A.~Tesi,
  ``Implications of a Light Higgs in Composite Models,''
  arXiv:1205.0232 [hep-ph].

\bibitem{marzocca}
  D.~Marzocca, M.~Serone and J.~Shu,
  ``General Composite Higgs Models,''
  JHEP {\bf 1208} (2012) 013
  [arXiv:1205.0770 [hep-ph]].

\bibitem{pomarol}
  A.~Pomarol and F.~Riva,
  ``The Composite Higgs and Light Resonance Connection,''
  JHEP {\bf 1208} (2012) 135
  [arXiv:1205.6434 [hep-ph]].

\bibitem{Bounds53_CMS}
 The CMS Collaboration,
 ``Search for a heavy partner of the top quark with charge 5/3'',
 CMS-PAS-B2G-12-003.

\bibitem{Bounds53_ATLAS}
 The ATLAS Collaboration,
 ``Search for exotic same-sign dilepton signatures (b' quark,
   $T_{5/3}$ and four top quarks production) in 4.7/fb of pp
   collisions at $\sqrt{s}=7$~TeV with the ATLAS detector'',
   ATLAS-CONF-2012-130.

\bibitem{Cohen:1996vb}
  A.~G.~Cohen, D.~B.~Kaplan and A.~E.~Nelson,
  ``The More minimal supersymmetric standard model,''
  Phys.\ Lett.\ B {\bf 388} (1996) 588
  [hep-ph/9607394].

\bibitem{AAAR}
 A.~De Simone, O.~Matsedonskyi, R.~Rattazzi and A.~Wulzer, to appear.

\bibitem{Panico:2007qd}
  G.~Panico and A.~Wulzer,
  ``Effective action and holography in 5D gauge theories,''
  JHEP {\bf 0705} (2007) 060
  [hep-th/0703287].


\end{thebibliography}
\end{document}